\documentclass[pra,twocolumn,showkeys,showpacs,superscriptaddress,amsmath,amsfonts]{revtex4-1}
\usepackage[utf8x]{inputenc}
\usepackage{microtype}
\usepackage{lmodern}
\usepackage{graphicx}
\usepackage{hyperref}
\usepackage{color}
\usepackage{subfigure}

\newcommand{\bra}[1]{\langle#1\rvert}
\newcommand{\ket}[1]{\lvert#1\rangle}
\newcommand{\mean}[1]{\langle#1\rangle}
\allowdisplaybreaks[1]

\begin{document}

\title{Green's function variational approach to orbital polarons
       in KCuF$_3$}

\author{Krzysztof Bieniasz}
\email{krzysztof.bieniasz@uj.edu.pl}
\affiliation{Marian Smoluchowski Institute of Physics, Jagiellonian University,
             prof. S. {\L}ojasiewicza 11, PL-30348 Krak\'ow, Poland}
\affiliation{Department of Physics and Astronomy, University of British Columbia,
             Vancouver, British Columbia, Canada V6T 1Z1}
\affiliation{\mbox{Stewart Blusson Quantum Matter Institute, University of British Columbia,
             Vancouver, British Columbia, Canada V6T 1Z4}}

\author{Mona Berciu}
\affiliation{Department of Physics and Astronomy, University of British Columbia,
             Vancouver, British Columbia, Canada V6T 1Z1}
\affiliation{\mbox{Stewart Blusson Quantum Matter Institute, University of British Columbia,
             Vancouver, British Columbia, Canada V6T 1Z4}}

\author{Maria Daghofer}
\affiliation{Institute for Functional Materials and Quantum Technologies, University of Stuttgart,\\ Pfaffenwaldring 57, D-70550 Stuttgart, Germany}

\author{Andrzej M. Ole\'s}
\affiliation{Marian Smoluchowski Institute of Physics, Jagiellonian University,
             prof. S. {\L}ojasiewicza 11, PL-30348 Krak\'ow, Poland}
\affiliation{Max Planck Institute for Solid State Research,
             Heisenbergstra{\ss}e 1, D-70569 Stuttgart, Germany}

\date{6 July 2016}

\begin{abstract}
We develop an $e_g$ orbital, $t$-$J$-like model of a single charge
doped into a two-dimensional plane with ferromagnetic spin order and
alternating orbital order, and present its solution by Green's
functions in the variational approximation framework. The model is
designed to represent the orbital physics within ferromagnetic $(a,b)$
planes of KCuF$_3$ and K$_2$CuF$_4$. The variational approximation
(VA) relies on the systematic generation of equations of motion for
the Green's function, taking into account the real-space constraints
coming from the exclusion of doubly occupied sites. This method is
compared to the firmly established self-consistent Born approximation,
and to the variational cluster approximation (VCA) which relies on the
itinerant regime of the model. We find that the present variational
approximation captures
the essential aspects of the spectral weight distribution of the coherent
quasiparticle state and gives a result similar to the VCA, while also
reproducing well the momentum dependence of the spectral moments.
In contrast, the spectral function obtained within the self-consistent
Born approximation is more incoherent and its quasiparticle is heavier,
at strong effective couplings, than observed with VCA and VA.
\end{abstract}


\maketitle

\section{Introduction}
\label{sec:intro}

Strongly correlated electron systems with active orbital degrees of
freedom pose a variety of challenging problems. One of them is charge
propagation in systems with orbital order which has a whole plethora
of possible processes involving the ordered background~\cite{Zaa93}.
Some of them could support coherent hole propagation, but in general
the charge is expected to be confined when the spin-orbital order is
robust \cite{Woh09}. This problem simplifies tremendously when the
orbital degrees of freedom are quenched, as in high $T_c$
superconducting cuprates, and hole propagation in the antiferromagnetic
(AF) background may be studied for a two-dimensional (2D) square
lattice~\cite{Lee06} within the effective $t$-$J$ model or its
extensions~\cite{Toh94,*Bal95}. In these systems individual CuO$_2$
planes are separated from one another and it is sufficient to
investigate charge propagation within a single plane. A hole added to
a quantum antiferromagnet propagates then by processes which involve
quantum fluctuations in the spin background and this propagation is
controlled by a new energy scale set by the superexchange
$J$~\cite{Mar91}. In contrast, a hole is almost blocked in the absence
of quantum fluctuations for Ising exchange interactions --- then the
hopping in a square lattice is possible only by self-repairing
processes on plaquettes, as recognized by Trugman~\cite{Tru88}.

In this context the orbital models are of particular interest as they
offer several different scenarios as the exchange interactions are
always of lower symmetry than SU(2) and also the orbital flavor is not
conserved in some cases. They originate from the spin-orbital physics
\cite{Kug82,Fei97,*Fei98,Fei99,Kha00,Kha01,*Kha04,Ole05,Kha05,Rei05,
Cuo06,Cha08,Zaa09,Cor12,Brz12,*Brz13,*Brz11,*Cza15,Woh11,*Mar12,*Bis15,
*Plo16,*Woh13,*Che15,Kumar,Mil14,Ave15,*Hor11,*Ave13,Kat15,Brz15,*Brzjs},
and arise when spin order is ferromagnetic (FM) --- then the
superexchange reduces to orbital interactions only and spin-orbital
entanglement \cite{Ole12,*Ole15} is absent in the ground state
\cite{Ole06,You15,*You12,*Hor15}. Realistic 2D or three-dimensional
(3D) $e_g$ orbital models are also strongly frustrated, but
nevertheless alternating orbital (AO) order arises at finite
temperature and is induced by the strongest interaction~\cite{Ryn10},
while quantum effects are small \cite{vdB99}.
Doping of the $e_g$ orbital system suppresses
gradually orbital order \cite{Tan05,*Tan09} and gives a disordered
orbital liquid which plays a role in doped FM manganites~\cite{Fei05}.
It was argued that orbital non-conserving processes (see below) are
responsible for the onset of the disordered orbital liquid for larger
doping. Indeed, such a disordered orbital state supports the FM
metallic phase in doped manganites and was invoked to explain the
observed cubic symmetry in the 3D magnon dispersion for
La$_{1-x}$Pb$_x$MnO$_3$ \cite{Kil00,*Ole02}.

Concerning the hole propagation, it is important to realize that the
orbital models are more classical than spin ones, for instance the
orbital model for $t_{2g}$ orbitals has Ising superexchange and thus
realizes the paradigm introduced by Trugman~\cite{Tru88}. One expects
that in absence of orbital fluctuations holes should be immobile. Yet,
weak hole propagation becomes possible due to three-site processes for
$t_{2g}$ orbitals, which appear in doped systems in the same order
of the perturbation theory as the superexchange itself
\cite{Dag08,*Woh08,Wro10}. The three-site terms dominate also hole
propagation in the 2D compass model~\cite{Brz14,*Dag15} which is
related to the $e_g$ orbital model \cite{Cin10}. Finally, recently it
was argued that the concept of hole-localization in the absence of
quantum fluctuations does not to apply to three-band models
\cite{Ebr14,Ebr15}.

Perhaps the most complex situation arises in the $e_g$ orbital model
where both the orbital superexchange interactions~\cite{vdB99} and
the kinetic energy~\cite{Fei05} are radically different from their
counterparts in the spin $t$-$J$ model. The interactions are Ising-like
in a one-dimensional (1D) model~\cite{Dag04} and involve one directional
($3z^2-r^2$)-like orbital and one perpendicular to it ($x^2-y^2$)-like
orbital, both selected by the bond direction. Therefore, the 1D orbital
model is classical but quantum fluctuations increase with increasing
dimension: the 2D model and even more so the 3D model become quantum,
but still both have much weaker quantum fluctuations~\cite{vdB99} than
the SU(2) symmetric Heisenberg spin model.

In addition, the hopping in dimension higher than one does not
conserve the orbital flavor and includes both the intraorbital processes
with and interorbital ones without orbital conservation~\cite{Fei05}.
The orbital-flip processes allow an alternative mode of hole
propagation, where the orbital order of the ground state is not
disturbed at all.  As the hole thus needs neither spin-flip nor
three-site terms to move, one would expect to observe a quasiparticle
(QP) dispersion on a scale of the (interorbital) hopping $t$ rather
than on the scale of $J$. The additional presence of orbital-flipping
hoppings mediates interactions with the background that can further
renormalize the bandwidth, however the extremely small dispersion
found with the self-consistent Born approximation (SCBA)~\cite{Mar91}
comes as a surprise and seems to contradict this picture~\cite{vdB00}.

The purpose of this paper is to present a systematic study of hole
propagation in the 2D $e_g$ orbital model. A full understanding of
this orbital-polaron is a necessary ingredient -- together with the
full understanding of the spin-polaron -- before one can hope to
understand the complex behavior of spin-orbital polarons, \emph{i.e.},
of the QPs that form in systems with strong coupling to
both orbital and spin excitations \cite{Woh09}.

To study the polaron of the 2D $e_g$ orbital model, we go beyond the
simplest SCBA approach and also investigate the consequences of the
local real space constraints on the hole propagation. By comparing with
the variational approach, we establish that the SCBA, which includes
only rainbow diagrams, gives a surprisingly good qualitative picture.
However, we also find that its one weakness is that it gives
predominantly incoherent spectra with too small QP bandwidth. Indeed,
we find that the other approaches yield a fairly robust dispersion, which
is more easily reconciled with the underlying propagation mechanism.
We also develop and validate a variational approach to the problem
that can more easily be generalized to the full 3D model.

The paper is organized as follows: In Sec.~\ref{sec:2D} we introduce
the $e_g$ orbital Hamiltonian for a 2D $(a,b)$ plane and identify the
processes responsible for free hole propagation as well as interactions
which will have to be treated by some approximate method. In this paper
we employ several methods introduced in Sec.~\ref{sec:method}:
(i) variational approach in Sec.~\ref{sec:var},
(ii) self-consistent Born approximation (SCBA) in Sec.~\ref{sec:scba},
(iii) variational cluster approximation (VCA) in Sec.~\ref{sec:vca}, and
(iv) spectral moment approach in Sec.~\ref{sec:mom}.
We then discuss the convergence of the variational approach upon
increasing the variational space, and present and compare with one
another the numerical results obtained using these various methods in
Sec.~\ref{sec:results}. The summary and conclusions are given in
Sec.~\ref{sec:summa}. The paper is completed by Appendix~\ref{sec:2orb}
which elaborates on the details of the variational solution.

\section{The 2D orbital model}
\label{sec:2D}

KCuF$_{3}$ is a 3D perovskite material, with staggered FM $(a,b)$
planes along the $c$ cubic axis \cite{Bella}. Each $(a,b)$ plane has
AO order following the Goodenough-Kanamori rules \cite{Goode,Kan59}.
A carrier doped into this material will couple to both spin and
orbital excitations to realize its motion \cite{Woh09}. As already
mentioned, our current goal is to understand the orbital-polaron that
forms when the carrier is restricted to move within a single $(a,b)$
plane. The spin degree of freedom is then quenched and only orbital
interactions survive.

To address this situation we start with the on-site Coulomb
interactions for two degenerate $e_g$ orbitals $\{z,\bar{z}\}$ at Cu
sites; they read \cite{Ole83},
\begin{align}
  \label{intra}
    H_{\mathrm{int}}&= U_0 \sum_{i,\mu=z,\bar{z}} n_{i\mu,\uparrow}
    n_{i\mu,\downarrow}\nonumber\\
    &+\sum_{i} \left(U_0-\frac{5}{2}J_{\mathrm{H}}\right)n_{iz}n_{i\bar{z}}
    -2J_{\mathrm{H}}\sum_{i}\mathbf{S}_{iz}\cdot\mathbf{S}_{i\bar{z}}\\
    &+ J_{\mathrm{H}}\sum_{i} \left( d^\dagger_{iz,\uparrow}
      d^\dagger_{iz,\downarrow} d_{i\bar{z},\downarrow}^{}
      d_{i\bar{z},\uparrow}^{}+ d^\dagger_{i\bar{z},\uparrow}
      d^\dagger_{i\bar{z},\downarrow} d_{iz,\downarrow}^{}
      d_{iz,\uparrow}^{}\right).\nonumber
\end{align}
The basis $\{z,\bar{z}\}$ was chosen as consisting of one directional
$z$ orbital along the cubic $c$ axis, and orthogonal to it $\bar{z}$
orbital within the $(a,b)$ plane as follows:
\begin{equation}
  \label{basis}
  \begin{split}
    \ket{z}&\equiv \ket{3z^2-r^2}/\sqrt{6},\\
    \ket{\bar{z}}&\equiv \ket{x^2-y^2}/\sqrt{2}.
  \end{split}
\end{equation}
The interactions in Eq.~\eqref{intra} are parametrized by two parameters:
intraorbital Coulomb element $U_0$ and Hund's exchange $J_{\mathrm{H}}$.
These parameters decide about the energies of the ground and excited
states. In the present case of Cu$^{2+}$ ions with $d^9$ configuration,
$H_{\mathrm{int}}$ does not contribute to the ground states of a single hole
at each site.

Among the excited states at Cu$^{3+}$ ions with $d^8$ ($t_{2g}^6e_g^2$)
configurations, the high-spin $S=1$ states, as for instance
$|z\!\uparrow\bar{z}\!\uparrow\rangle$, have the lowest energy of $(U_0-3J_H)$
\cite{Ole05}. These states with two holes occupying two orthogonal
orbitals of $d^8$ ions arise by charge excitations on the bonds within
the $(a,b)$ planes of KCuF$_3$ and decide about the FM order. Thus, as
long as we limit ourselves to a single FM plane, we can consider
spinless $d^{\dagger}_{i\mu,\uparrow}\equiv d^{\dagger}_{i\mu}$ fermions
and replace Eq.~\eqref{intra} with
\begin{equation}
  \label{eq:hub}
  \mathcal{H}_{U} = U \sum_{i} n_{iz} n_{i\bar{z}},
\end{equation}
where the effective interaction between two holes with different orbital
flavors at the same site $i$ is,
\begin{equation}
  \label{eq:U}
  U \equiv U_0-3J_H.
\end{equation}
Thus, from now on the effective parameter $U$ stands for the
interaction $(U_0-3J_H)$ which is the only interaction left in the
subspace of high-spin (triplet) excited states \cite{vdB99}.

We supplement the above interaction term with the kinetic energy for
$e_g$ spinless holes on the full 3D lattice assuming FM order~\cite{Fei05}:
\begin{equation}
  \label{eq:3d}
  \mathcal{H}_{t} = -t\sum_{\alpha} \sum_{\mean{ij} \| \alpha}
  \left( d_{iz_{\alpha}}^{\dag} d_{jz_{\alpha}}^{} +\mathrm{H.c.}\right),
\end{equation}
which describes the hopping between nearest neighbor (NN) sites along
each bond $\mean{ij}$ that involves two $\sigma$-bonding
$\ket{3z_{\alpha}^{2}-r^{2}}$-type orbitals oriented along the bond
direction $\alpha=\{a,b,c\}$, with $z_c\equiv z$, \emph{etc.}
As the hopping $t$ follows from a two-step process via ligand F($2p$)
orbitals, the $\delta$-hopping between two
$\ket{x_{\alpha}^{2}-y_{\alpha}^{2}}$-type orbitals, which
are perpendicular to the direction $\alpha$, vanishes by symmetry. We
use again the same short-hand notation with $x_c\equiv x$, $y_c\equiv y$,
\emph{etc.} The above formulation, while concise, is not very practical due
to the orbital basis changing depending on the $\alpha$ direction.
By transforming the Hamiltonian~\eqref{eq:3d} to the standard orthogonal
orbital basis \eqref{basis} one obtains:
\begin{multline}
  \label{eq:trans3d}
  \mathcal{H}_{t} = -t \sum_{\mean{ij} \| c} d_{iz}^{\dag} d_{jz}^{} \\
  -\frac{t}{4}\sum_{\mean{ij} \| a} \left[(d_{iz}^{\dag} -\sqrt{3} d_{i\bar{z}}^{\dag})
  (d_{jz}^{} -\sqrt{3} d_{j\bar{z}}^{})+\mathrm{H.c.}\right] \\
  -\frac{t}{4}\sum_{\mean{ij} \| b} \left[(d_{iz}^{\dag} +\sqrt{3} d_{i\bar{z}}^{\dag})
  (d_{jz}^{} +\sqrt{3} d_{j\bar{z}}^{}) + \mathrm{H.c.}\right].
\end{multline}
The results presented below are in units of $t=1$.

Here we are interested both in the orbital Hubbard model for $e_g$
electrons/holes \cite{Fei05} given by Eqs.~\eqref{eq:hub} and
\eqref{eq:trans3d}, and in effective large-$U$ models, \emph{i.e.},
superexchange models derived by second-order perturbation expansion of
the Hamiltonian. Following a procedure analogous to the $t$-$J$ model
derivation \cite{Cha77}, one can find the exchange interaction, which
in this case consists of four terms expressed using singlet/triplet
projection operators in spin and orbital Hilbert spaces.

Although in KCuF$_3$ the spin interactions in $(a,b)$ planes are rather
weak, the ground state has $A$-type AF spin and $C$-type
AO order~\cite{Pao02,*Bella}, \emph{i.e.}, we
can view the system as consisting of strongly coupled $(a,b)$ planes
with AO order that are stacked with ferro-orbital (FO) order along the
$c$ axis. AF correlations between the layers strongly suppress
hole motion in $c$ direction, making it feasible to focus on the
in-plane orbital physics separately from the system's behavior along
the $c$-direction, at least as long as we are considering a
system near its ground state.

To find the exchange Hamiltonian for a single 2D plane we need to
project out the spin-singlet states from the full Hamiltonian.
After averaging over the in-plane spin-triplet states one arrives
at~\cite{vdB99}:
\begin{equation}
  \label{eq:Hj}
  \mathcal{H}_{J} = \frac{J}{2} \sum_{\mean{ij}} [
  T_{i}^{z}T_{j}^{z} + 3 T_{i}^{x} T_{j}^{x}
  \mp \sqrt{3} (T_{i}^{x}T_{j}^{z}+T_{i}^{z}T_{j}^{x}) ],
\end{equation}
where $J=t^2/U$, the upper/lower sign is for the $\mean{ij}$ bond
oriented along $a/b$ axis respectively, and $\{T_i^{\alpha}\}$
operators are the same as spin operators (1/2 times the respective
Pauli matrix), only acting in the orbital space in which $z/\bar{z}$
states correspond to spin up/down states.  Alternatively, one may
derive Eq.~\eqref{eq:Hj} from the Kugel-Khomskii model
\cite{Kug82,Fei97} by considering a FM system with superexchange
given by triplet charge excitations
$d_i^9d_j^9\leftrightarrow d_i^8d_j^{10}$.

\begin{figure}[t!]
  \centering
  \includegraphics[width=0.65\columnwidth]{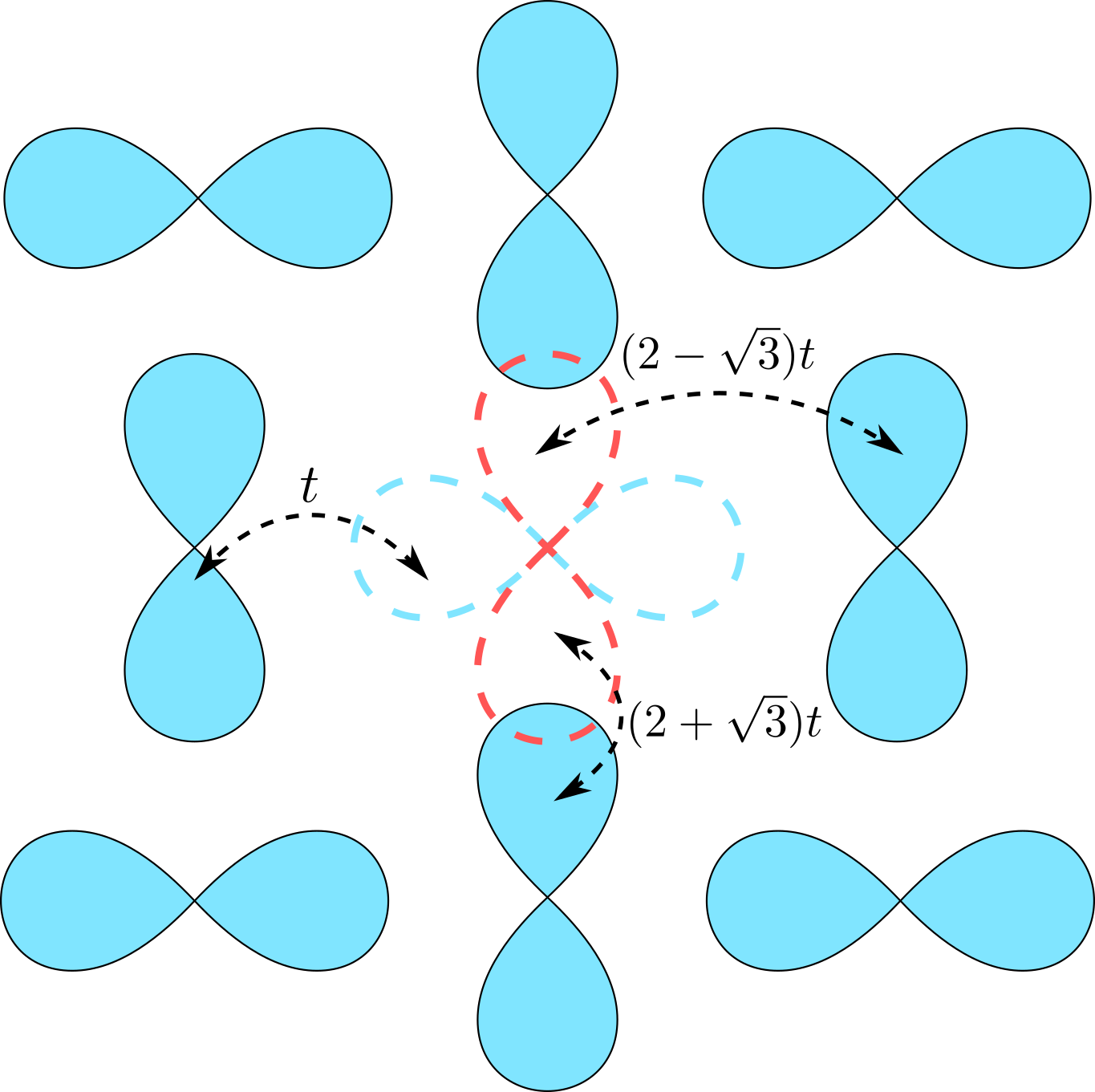}
\caption{(Color online)
Schematic representation of the kinetic Hamiltonian $\mathcal{H}_t$
\eqref{eq:trans3d} and the ground state with AO order of the 2D orbital
model $\mathcal{H}_{J}$ \eqref{eq:Hj}. The
bow-shapes represent the $\ket{\pm}$ states which are highly directional
in the $(a,b)$ plane. Full orbitals represent a singly occupied state
(corresponding to the overall $d^9$ state) and the blue color represents
the orbitals occupied in the ground state. The dashed orbital contours
represent a doubly occupied central site (\emph{i.e.}, electron doping).
Blue contour refers to the orbital normally half-filled in the ground
state and the red one indicates the orbital that would have been wrongly
occupied if an electron were to be removed from it. The kinetic elements
of Hamiltonian~\eqref{eq:transfH} are a consequence of orbital symmetry
and overlap.}
\label{fig:model}
\end{figure}

For an FM plane in KCuF$_3$, or for the 2D compound K$_2$CuF$_4$, it has
been well established through a variety of analytical and numerical
treatments \cite{vdB99,Cin10,vdB04,Dag06} that the leading superexchange
term $3T^x_iT^x_j$ dominates Eq.~\eqref{eq:Hj} and stabilizes the AO
order shown schematically in Fig.~\ref{fig:model}. Indeed,
the orbital ground state has been found to be composed of alternating
eigenstates of $T^x$ operators, \emph{i.e.},
\begin{equation}
\ket{\pm}=\frac{1}{\sqrt{2}}(\ket{z}\pm\ket{\bar{z}}).
\label{eq:pm}
\end{equation}
Orbital excitations from such an ordered ground state have been studied
by several authors in the past
\cite{vdB99,Woh11,*Mar12,*Bis15,*Plo16,*Woh13,*Che15,Oka02,Ish05,For08,Nas13}.
Below we use them to investigate orbital polarons when a hole is added.

Quantum fluctuations around perfect N\'eel order involve one- and
two-orbiton (orbital excitation) processes and have consistently been
found to be small. As a further check, we carried out the SCBA with
and without these quantum fluctuations and have found the results to
be almost identical. Thus, while the other
two terms can be included in the treatment (they are automatically
included in the VCA) in the following we restrict the variational
approach to only the Ising-exchange term $ \mathcal{H}_{J} \approx
\frac{3}{2}J \sum_{\mean{ij}} T^x_iT^x_j$, for the sake of simplicity.

The kinetic Hamiltonian~\eqref{eq:trans3d}, after being restricted to
the $(a,b)$ plane, is transformed to the basis~\eqref{eq:pm}. Next,
the orbital degree of freedom is decoupled from the fermionic
operators by means of the slave boson  formalism~\cite{Mar91}:
\begin{equation}
  \label{eq:slave}
  d_{i0}^{\dag} = f_{i}^{\dag}, \quad d_{i1}^{\dag} = f_{i}^{\dag} a_{i}^{},
\end{equation}
where the $\{0,1\}$ indices on the left hand side indicate fermion
creation in the ground or excited state, respectively, \emph{i.e.},
$\ket{+}$ or $\ket{-}$~\eqref{eq:pm}, depending on the sublattice since
the orbital state is alternating. This transformation works under the
assumption that $f_{i}^{\dag}: d^9\to d^{10}$ and $f_{i}: d^{10}\to d^9$,
which is the consequence of disallowing doubly occupied states
in the Hilbert space.

After performing all the above transformations, we arrive at the final
form of the strong coupling Hamiltonian,
\begin{equation}
\mathcal{H}=\mathcal{H}_{J}+\mathcal{T}+\mathcal{V} ,
\end{equation}
where:
\begin{subequations}
  \label{eq:transfH}
  \begin{align}
    \label{eq:transfHJ}
    \mathcal{H}_{J} &= \frac{3}{8}J\sum_{\mean{ij}} (1 - \sigma_{i}^{z}\sigma_{j}^{z}),\\
    \label{eq:transfT}
    \mathcal{T} &= -\frac{t}{4}\sum_{\mean{ij}} (f_{i}^{\dag}f_{j}^{} + \mathrm{H.c.}) =
    \sum_{\mathbf{k}} \epsilon_{\mathbf{k}}^{} f_{\mathbf{k}}^{\dag} f_{\mathbf{k}}^{},\\
    \label{eq:transfV}
    \mathcal{V} &= -\frac{t}{4} \sum_{i,\delta} \left[
      (2+\sqrt{3} e^{i\pi_{y}\cdot\delta}
      e^{i \mathbf{Q}\cdot\mathbf{R}_{i}}) a_{i}^{\dag} +\right.\nonumber\\
    &\left.+(2-\sqrt{3} e^{i\pi_{y}\cdot\delta}
      e^{i \mathbf{Q}\cdot\mathbf{R}_{i}})
      a_{i+\delta}^{} + a_{i}^{\dag} a_{i+\delta}^{} \right]
    f_{i+\delta}^{\dag} f_{i}^{}.
  \end{align}
\end{subequations}
Here, $\epsilon_{\mathbf{k}}^{} = -t\gamma_{\mathbf{k}}^{}$ is the free
particle dispersion, with
$\gamma_{\mathbf{k}}^{}=(1/z)\sum_{\delta} e^{i\mathbf{k}\cdot\delta}$.
Also, for simplicity $\mathcal{H}_{J}$ and the corresponding ground
state have been transformed from AO order into the FO one, and the
leading Ising term has the exchange constant $J$ with an explicit
factor of $-3/8$. Note that we also subtracted the background energy
out of $ \mathcal{H}_{J}$. The free propagation $\mathcal{T}$
\eqref{eq:transfT} is the consequence of the kinetic energy terms
conserving the orbital flavor, while the interaction $\mathcal{V}$
\eqref{eq:transfV} arises by the same mechanism as the hole-magnon
coupling in the spin model \cite{Mar91}.
The phase factors serve to accommodate the model's dependence on
direction: $\pi_y=(0,\pi)$, $\mathbf{Q}=(\pi,\pi)$, and $\delta$ is
a vector variable pointing to the nearest neighbors of site~$i$.

\section{Methods}
\label{sec:method}

\subsection{Variational approximation (VA)}
\label{sec:var}
Starting from the Bloch state of a free electron,
\begin{equation}
\ket{\mathbf{k}}=f_{\mathbf{k}}^{\dag}\ket{0}=
\frac{1}{\sqrt{N}}\sum_{i}e^{i\mathbf{k}\cdot\mathbf{R}_i}f_{i}^{\dag}\ket{0},
\label{eq:kstate}
\end{equation}
we define the Green's function as
\begin{equation}
G(\mathbf{k},\omega)\equiv\bra{\mathbf{k}}\mathcal{G}(\omega)\ket{\mathbf{k}},
\end{equation}
where $\mathcal{G}(\omega)=(\omega+i\eta-\mathcal{H})^{-1}$  is the
resolvent. Here, $|0\rangle$ is the N\'eel AO ground-state of the
undoped 2D plane,  described above.

The variational approximation is based on a simplification of the
equations of motion which are obtained through repeated Dyson
expansions of the resolvent:
\begin{equation}
  \label{eq:dyson}
  \mathcal{G}(\omega) = \mathcal{G}_{0}(\omega)
  +\mathcal{G}(\omega)\mathcal{V}\mathcal{G}_{0}(\omega).
\end{equation}
Here,
$\mathcal{H}_{0} = \mathcal{T}+\mathcal{H}_{J}$ corresponds to
$\mathcal{G}_{0}(\omega)$ and $\mathcal{V}$ is given by Eq.~\eqref{eq:transfV}.

The rationale underlying the variational approach is that the creation
of every orbiton costs an energy proportional to $J$, so the bigger
the value of $J$ the less orbitons are likely to appear in the QP
cloud in the ground state. Therefore, one can restrict the Dyson
expansion in terms of the number of orbitons generated in the system,
and one can further decrease the Hilbert space by constraining the
orbitons to be located close to each other. This approach can of
course be tested by verifying the convergence of the result as a
function of the number of orbitons allowed in the system, \emph{i.e.},
by increasing the variational space. In this Section we show how this
approach works in the framework of a one orbiton approximation, when
only up to one orbiton is allowed to appear in the QP cloud. Then we
briefly discuss the generalization for the two-orbiton case, which is
further detailed in the Appendix. Note that we have implemented this
variational approach, and show results,  for  up to six orbitons in
the QP cloud. However, those equations of motion become too cumbersome
to write explicitly. We also note that similar variational schemes
have proved to be very accurate for the study of various models of
lattice polarons \cite{Ber06,Goo06,Ber07,Mar10} as well as spin
polarons \cite{Ber11,Tro13,Ebr14,Ebr15}.

Using the Dyson expansion and evaluating $\mathcal{V}\ket{\mathbf{k}}$
explicitly in real space, we arrive at the following expression:
\begin{equation}
  \label{eq:green}
  \begin{split}
&G(\mathbf{k},\omega) = \left[ 1-\frac{t}{2}\sum_{\delta}
F_{1}(\mathbf{k},\omega,\delta)+\right.\\
&\left.-\frac{\sqrt{3}t}{4}\sum_{\delta} \bar{F}_{1}(\mathbf{k},\omega,\delta)
e^{i\mathbf{\pi}_y\cdot\mathbf{\delta}} \right] G_{0}(\mathbf{k},\omega-4J'),
  \end{split}
\end{equation}
where
$G_0((\mathbf{k},\omega)=(\omega-\epsilon_{\mathbf{k}}+i\eta)^{-1}$ is
the free-electron Green's function, the shorthand notation $J'=3J/8$
has been introduced for convenience, and we define:
\begin{align}
  \label{eq:f1}
F_{1}(\mathbf{k},\omega,\delta) &=
\bra{\mathbf{k}} \mathcal{G}(\omega) \frac{1}{\sqrt{N}}
\sum_{i} e^{i\mathbf{k}\cdot\mathbf{R}_{i}}
f_{i+\delta}^{\dag} a_{i}^{\dag} \ket{0},\\
\label{eq:f1b}
\bar{F}_{1}(\mathbf{k},\omega,\delta) &= \bra{\mathbf{k}}
\mathcal{G}(\omega)
\frac{1}{\sqrt{N}} \sum_{i} e^{i(\mathbf{k+Q})\cdot\mathbf{R}_{i}}
f_{i+\delta}^{\dag} a_{i}^{\dag} \ket{0},
\end{align}
as the generalized Green's functions for the state with a single
orbital excitation above the ground state.

In order to find these new propagators, we use the Dyson expansion
again to generate their equations of motion. In the one-orbiton
variational scheme we include only the
part of $\mathcal{V}$ which does not create new orbitons.
However, at this point one has to impose additional constraints on the
movement of the electron, namely: ($i$) the electron and the orbiton
cannot both occupy the same site, and ($ii$) when the electron and
orbiton are far apart the energy increase is by $12J'$ (4 broken bonds
accompanied by 4 excited bonds), however if they are on adjacent sites
the energy increase is only by $10J'$ (one excited bond less). The
above constraints can be taken into account by adding the following
terms to the $\mathcal{H}_{0}$ Hamiltonian, in
order to cancel the corresponding processes:
\begin{equation}
  \label{eq:vi}
  \mathcal{V}_{1} = \frac{t}{4} \sum_{\epsilon}
  (f_{i}^{\dag} f_{i+\epsilon}^{} + \mathrm{H.c.})
  -2J'\sum_{\epsilon} n_{i+\epsilon},
\end{equation}
where $i$ is the location of the orbiton and $n_{i+\epsilon}$ is
the electron number operator for the sites NN to site $i$. The above
expression modifies the free part of the Hamiltonian, which means
now instead of
$\mathcal{G}_{0}(\omega)$ one has to use $\mathcal{G}_{1}(\omega)$
(corresponding to $\mathcal{H}_{0}+\mathcal{V}_{1}$) in the Dyson
expansion of the generalized Green's function:
\begin{equation}
\mathcal{G}(\omega) = [1+\mathcal{G}(\omega)\mathcal{V}]
\mathcal{G}_{1}(\omega).
\end{equation}
However, $\mathcal{G}_{1}(\omega)$ is no longer diagonal in Fourier
space, and so its propagator needs to be solved first. This can again
be done using the Dyson equation,
\begin{equation}
\mathcal{G}_{1}(\omega) = [1+\mathcal{G}_{1}(\omega)\mathcal{V}_{1}]
\mathcal{G}_{0}(\omega),
\end{equation}
which leads to the following set of equations of motion:
\begin{equation}
  \label{eq:gi}
  \begin{split}
&G_{1}(n,i\!+\!\delta,\omega) = G_{0}(n,i\!+\!\delta,\omega-12J')
+\!\sum_{\epsilon} G_{1}(n,i\!+\!\epsilon,\omega)\\
    &\times\left[\frac{t}{4}G_{0}(i,i\!+\!\delta,\omega-12J')
    -2J'G_{0}(i\!+\!\epsilon,i\!+\!\delta,\omega-12J')\right],
  \end{split}
\end{equation}
where $G_{1}(n,i\!+\!\delta,\omega)=
\langle 0|f_n\mathcal{G}_1(\omega) f^{\dagger}_{i+\delta}|0\rangle$
describes the propagation of an
electron in real space, in the presence of the forbidden site $i$ (due
to the orbiton located there) from site  $\ket{i+\delta}$ to any site
$n\ne i$. However, as we show below, the only  propagators needed are
for NN sites $n=i+\gamma$. This leads to a reduction of
Eq.~\eqref{eq:gi} to a $4\times4$ matrix equation for the set of
functions $G_{1}(i\!+\!\gamma,i\!+\!\delta,\omega)$, describing
propagators exclusively around the orbiton excitation:
\begin{equation}
  \label{eq:gimatrix}
\mathbb{G}_{1}^{\gamma\delta} = \mathbb{G}_{0}^{\gamma\epsilon}
\left[\mathbb{I}^{\delta\epsilon} -\tfrac{t}{4}\mathbb{G}_{0}^{\delta 0}
+2J'\mathbb{G}_{0}^{\delta\epsilon}\right]^{-1}.
\end{equation}

What makes this method of including the site occupation constraints
particularly appealing, in spite of its complicated nature, is the
fact that the real space Green's function depends only
on the distance between sites:
\begin{equation}
  \label{eq:G0}
  G_{0}(m,n,\omega) = \frac{1}{\pi^{2}} \int_{0}^{\pi}d^{2}\mathbf{k}\,
  G_{0}(\mathbf{k},\omega)\,e^{i\mathbf{k}\cdot(\mathbf{R}_m-\mathbf{R}_n)},
\end{equation}
and not the specific location of the $\{\ket{m},\ket{n}\}$ states.
Moreover, it can be shown (e.g. see the approach introduced by Morita
\cite{Mor71a,*Mor71}) that for 2D lattice Green's functions, the
contributions $G_{0}(m,n,\omega)$ can be expressed exactly in terms of
the complete elliptic integrals $K(\kappa),E(\kappa)$, where
$\kappa=t/\omega$. Therefore, this method in principle allows for exact
analytical calculation of both the free as well as the constrained
non-interacting real space Green's functions.

Going back to the calculation of the $F_{1}$ functions, one has
to use the Dyson equation as explained above, which leads to:
\begin{widetext}
  \begin{align}
    \label{eq:f1s}
    F_{1}(\mathbf{k},\omega,\delta) &= -\frac{t}{4}\sum_{\gamma} \left[
      2G(\mathbf{k},\omega)+\sqrt{3}\bar{G}(\mathbf{k},\omega)e^{i\pi_{y}\cdot\gamma}
      + F_{1}(\mathbf{k},\omega,-\gamma)e^{-i\mathbf{k}\cdot\mathbf{\gamma}}\right]
    G_{1}(i+\gamma,i+\delta,\omega),\\
    \label{eq:f1bs}
    \bar{F}_{1}(\mathbf{k},\omega,\delta) &= -\frac{t}{4}\sum_{\gamma}
    \left[2\bar{G}(\mathbf{k},\omega)+\sqrt{3}G(\mathbf{k},\omega)
      e^{i\mathbf{\pi}_y\cdot\mathbf{\gamma}} -\bar{F}_{1}(\mathbf{k},\omega,-\gamma)
      e^{-i\mathbf{k}\cdot\mathbf{\gamma}}\right] G_{1}(i+\gamma,i+\delta,\omega),
\end{align}
where
\begin{equation}
    \label{eq:gb}
\bar{G}(\mathbf{k},\omega) = \bra{\mathbf{k}}\mathcal{G}(\omega)\ket{\mathbf{k}+\mathbf{Q}} =
-\frac{t}{4}\sum_{\delta} \left[2\bar{F}_{1}(\mathbf{k},\omega,\delta)
+\sqrt{3}F_{1}(\mathbf{k},\omega,\delta)e^{i\mathbf{\pi}_y\cdot\mathbf{\delta}}\right]
G_{0}(\mathbf{k}+\mathbf{Q},\omega-4J').
\end{equation}
\end{widetext}
Altogether, Eqs.~\eqref{eq:green},~\eqref{eq:gimatrix}, and
Eqs.~\eqref{eq:f1s} to~\eqref{eq:gb} form a set of 14 coupled equations.
Solving this system numerically yields the desired Green's function
$G(\mathbf{k},\omega)$ for an electron moving in an AO system with at
most a single orbiton.

Following a procedure similar to the one outlined above, one can
perform a multi-orbiton approximation. For example, for a two-orbiton
variational space, one needs to perform
Dyson expansion three times, disallowing the three-orbiton states in
the last expansion. This leads to the creation of a series of
$F_{2}(\mathbf{k},\omega,\epsilon,\gamma)$ generalized Green's
functions. Here $\{\epsilon,\gamma\}$ are unit vectors setting the
path of the moving charge, leaving behind a trail of two orbitons,
described by
$\sum_i e^{i\mathbf{k}\cdot\mathbf{R}_i} f_{i+\epsilon+\gamma}^{\dag}a_{i+\epsilon}^{\dag}a_{i}^{\dag}\ket{0}$.
The set of equations is now closed  by tying
$F_{2}$ only to other $F_{2}$ functions and
$F_{1}$ functions defined before. This system of equations can
be easily solved numerically, similarly to the previous case.

For two-orbiton states there are, of course, two
no-double-occupancy constraints, and therefore now one has to
calculate new constrained ``free'' Green's functions, denoted
$G_{2}(m,n,\omega)$. Their matrix equation turns out to be very
similar to the one-orbiton case~\eqref{eq:gimatrix}, where all the
summations over nearest neighbor vectors have to be replaced by
summation over all the sites adjacent to the string of orbitons
present in the system.

A more detailed discussion of this two-orbiton solution is found in
the Appendix. The generalization to more orbitons
follows along similar lines. While in principle the relative distance
between the orbitons can be arbitrarily large, we expect that the
configurations with the highest weight in the QP cloud are those with
orbitons located in close vicinity to each other. We therefore impose
the relatively liberal restriction that any two orbitons have to be
within a distance not greater than the total number of orbitons in the
current state. This allows for boson clouds that are spatially
constrained, yet are not limited to unbroken strings.

\subsection{Self-consistent Born approximation (SCBA)}
\label{sec:scba}

Alternatively, the model can be solved by means of the SCBA, a method
well established in polaron research~\cite{Mar91,Ram98,vdB00}, which
has been proven to be highly accurate for spin-polarons in the $t$-$J$
model \cite{DMC,Konrad} but very poor for lattice polarons at strong
couplings, \emph{e.g.} in the Holstein model \cite{Ber06,Goo06}. It is
therefore not \emph{a priori} clear how accurate it is for orbiton
polarons. SCBA has been discussed extensively in the literature so here
we will limit ourselves to stating its form for the specifics of our
model.

SCBA is a self-consistent method based on the calculation of the
self-energy $\Sigma(\mathbf{k},\omega)$, assuming that all crossing
diagrams are negligibly small. In order to use it, we first need to
transform the Hamiltonian into its momentum representation using a
discrete Fourier transform. Furthermore,
$\mathcal{H}_J$ has to be simplified to linear orbital wave (LOW) order
by representing it with a quadratic form in bosonic orbiton operators
$\{a^{\dag},a\}$. Orbital fluctuations can be included at little extra
cost only requiring an additional step of the Bogoliubov transformation.

Our Hamiltonian transformed to Fourier space takes the form:
\begin{align}
  \label{eq:fouH}
\mathcal{V} &= \frac{-t}{\sqrt{N}} \sum_{\mathbf{k}\mathbf{q}}
\left[f_{\mathbf{k}}^{\dag} f_{\mathbf{k}-\mathbf{q}}^{}
\left( M_{\mathbf{k}\mathbf{q}}^{} \alpha_{\mathbf{q}}^{}
+ N_{\mathbf{k}\mathbf{q}}^{} \alpha_{\mathbf{q}+\mathbf{Q}}^{}
\right) + \mathrm{H.c.}\right],\\
\mathcal{H}_{J} &= \sum_{\mathbf{q}} \omega_{\mathbf{q}}^{}
\alpha_{\mathbf{q}}^{\dag}\alpha_{\mathbf{q}}^{},
\end{align}
where $\alpha_{\mathbf{q}}^{(\dag)}$ is the Bogoliubov-transformed
orbiton operator, $\{u_{\mathbf{q}},v_{\mathbf{q}}\}$ are the Bogoliubov
coefficients,
$\eta_{\mathbf{q}}=\gamma_{\mathbf{q}+\mathbf{\pi}_y}$, and
\begin{equation}
\omega_{\mathbf{q}}=3J\sqrt{1+\gamma_{\mathbf{q}}/3}\approx 3J
\end{equation}
is the orbiton energy (with and without fluctuations, respectively).
The carrier-orbiton interaction gives two vertex functions,
\begin{align}
M_{\mathbf{k}\mathbf{q}}&=2\,\left(u_{\mathbf{q}}
\gamma_{\mathbf{k}-\mathbf{q}}+v_{\mathbf{q}}\gamma_{\mathbf{k}}\right), \\
N_{\mathbf{k}\mathbf{q}}&=\sqrt{3}\,\left(u_{\mathbf{q}+\mathbf{Q}}
\eta_{\mathbf{k}-\mathbf{q}}+v_{\mathbf{q}+\mathbf{Q}}\eta_{\mathbf{k}}\right).
\end{align}
Having written the Hamiltonian in
this form, we can calculate the self-energy from:
\begin{multline}
  \label{eq:scbaself}
\Sigma(\mathbf{k},\omega) = \frac{t^{2}}{N}\sum_{\mathbf{q}}
\left[M_{\mathbf{k}\mathbf{q}} G(\mathbf{k}-\mathbf{q},\omega
  -\omega_{\mathbf{q}}) M_{\mathbf{k}\mathbf{q}}\right.\\
\left.+N_{\mathbf{k}\mathbf{q}} G(\mathbf{k}-\mathbf{q},\omega
-\omega_{\mathbf{q}+\mathbf{Q}}) N_{\mathbf{k}\mathbf{q}}\right],
\end{multline}
where $G$ is the SCBA Green's function, which is \emph{a priori}
unknown. Therefore, in the self-consistent calculation we start with
the free Green's function $G_{0}(\mathbf{k},\omega)=1/\omega$. Having
calculated the first order approximation of the self-energy, we can
find the next order approximation of the Green's function with
\begin{equation}
  \label{eq:scbagreen}
G(\mathbf{k},\omega) = \left[G_{0}^{-1}(\mathbf{k},\omega)
-\Sigma(\mathbf{k},\omega)\right]^{-1}.
\end{equation}
This solution can then be plugged back into Eq.~\eqref{eq:scbaself}
to obtain a better approximation of the self-energy, and so on, until
one reaches a stable result. This procedure usually converges very
fast, often finding a stable solution after as few as one or two steps.

As already stated, the SCBA only sums non-crossing (rainbow)
diagrams. It therefore ignores contributions from processes where the
electron absorbs the orbitons in a different order than their reversed
creation order. It also fails to impose the constraints of at most one
orbiton at one site, and of not allowing the electron and an orbiton on
the same site. As such, SCBA is expected to be less accurate than the
variational method which implements these local constraints exactly and
also includes the crossed diagrams (in its multi-orbiton flavors).

\subsection{Variational cluster approximation (VCA)}
\label{sec:vca}

We complement the approaches discussed above with a numerical treatment
based on the self-energy functional approach, the VCA. In this scheme,
the one-particle spectral density of a large system (e.g. in the
thermodynamic limit) is obtained by approximating its self-energy by
the self-energy of a small cluster~\cite{Pot03}. We do this here
numerically using the Lanczos algorithm to solve the Hamiltonian on
eight-, ten- and twelve-site clusters. Electronic correlations and
quantum fluctuations on short length scales within the cluster are
thus taken into account.

Long-range order is also included on a mean-field level via an
optimization of the grand potential with respect to the order parameter
for orbital order~\cite{Dah04}. The grand potential is in turn again
obtained from:
(i) the grand potential of the small cluster, and
(ii) the Green's function of the large system.
The wave function of the ordered orbital can likewise be optimized;
this procedure reproduces the expected result that the $\ket{\pm}$
orbital combinations~\eqref{eq:pm} alternate in a checkerboard pattern,
as depicted in Fig.~\ref{fig:model}. The ``optimal'' state
corresponding to a stationary point of the grand potential can then be
used to evaluate the one-particle spectral density for comparison with
the other approaches.

Unfortunately, the self-energy approach is only valid for interactions
that are contained purely within the directly solved cluster. In our
case, this implies that we cannot address the $t$-$J$ Hamiltonian
comprised of Eqs.~\eqref{eq:3d} and~\eqref{eq:Hj}, but instead have to
work on the itinerant Hubbard variant given by Eqs.~\eqref{eq:3d} and
\eqref{eq:hub}. In order to meaningfully compare the orbital Hubbard
model to the $t$-$J$-model results, we then have to restrict ourselves
to the regime of large $U/t$, \emph{i.e.}, small $J/t$. The physics of
this regime, however, can be expected to be well described if the
impact of quantum fluctuations can be considered short-range enough to
be captured by the directly solved cluster. As we find that results of
clusters with eight, ten, and twelve sites agree, we assume that
finite-size effects are not too severe.

\subsection{Spectral moments}
\label{sec:mom}

In order to gauge the accuracy of SCBA and the variational approach,
it is useful to calculate their numerical spectral moments and compare
them against an analytical calculation. The spectral moments are defined
as
\begin{equation}
\label{eq:moment}
M_{n}(\mathbf{k})= \int\limits_{-\infty}^{\infty}
A(\mathbf{k},\omega)\,\omega^n\,d\omega,
\end{equation}
where
\begin{equation}
\label{eq:ak}
A(\mathbf{k},\omega)=-\frac{1}{\pi}\,\Im[G(\mathbf{k},\omega)],
\end{equation}
is the normalized spectral function. Eq.~\eqref{eq:moment} is useful
for numerical integration of a calculated spectral function
$A(\mathbf{k},\omega)$. On the other hand, the analytical expression for
a spectral moment can be calculated directly from the model Hamiltonian,
using the algebraic formula~\cite{Nol72}:
\begin{equation}
  \label{eq:moment2}
  M_{n}(\mathbf{k}) =
  \bra{0} \big\{
  \underbrace{[[[f_{\mathbf{k}}^{},H],H],\ldots]}_{n-p},
  \underbrace{[\ldots,[H,[H,f_{\mathbf{k}}^{\dag}]]]}_{p}
  \big\} \ket{0},
\end{equation}
where the value of $0\le p\le n$ is arbitrary, \emph{i.e.}, it can be
chosen in the most convenient way for the problem at hand. Obviously,
$M_{0}(\mathbf{k})=1$ is just the spectral function normalization,
which is consistent with the integral of~\eqref{eq:ak}. Several higher
order moments were calculated explicitly:
\begin{align}
  \label{eq:mom}
  M_{1}(\mathbf{k}) &= \epsilon_{\mathbf{k}} +4J',\\
  M_{2}(\mathbf{k}) &=(\epsilon_{\mathbf{k}} +4J')^{2} +\tfrac{7}{4}t^{2},\\
  M_{3}(\mathbf{k}) &=(\epsilon_{\mathbf{k}} +4J')^{3}
  +\tfrac{57}{16}t^{2}\epsilon_{\mathbf{k}} +\tfrac{63}{2}t^{2}J'.
\end{align}
$M_{4}(\mathbf{k})$ was also calculated and is shown below, however its
expression is too long to be written here explicitly.

\section{Results and discussion}
\label{sec:results}

\subsection{Convergence of the variational approach}
\label{sec:convergence}

We start by analyzing the convergence of the spectral weight predicted
by the variational approach, with the increase of the variational
space (maximum number $n$ of orbitons allowed). Note that we always plot
$\tanh[A(\mathbf{k},\omega)]$ to highlight the low amplitude part of
the spectra. This transformation has the advantage that the relative
amplitudes of values smaller than 1 are nearly unaffected
(since $\tanh(x)\approx x-x^{3}/3+\ldots$ is nearly linear),
while the big values are all treated uniformly, because of the upper
limit of 1. Therefore, density maps presented herein indicate actual
maxima in black, while the incoherent parts of the spectral function
are shown in shades of gray. The dashed red lines serve as reference,
and mark the location of the free electron band at
$\omega=\epsilon_{\mathbf{k}}+4J'$.

\begin{figure}[t!]
  \centering
  \includegraphics[width=\columnwidth]{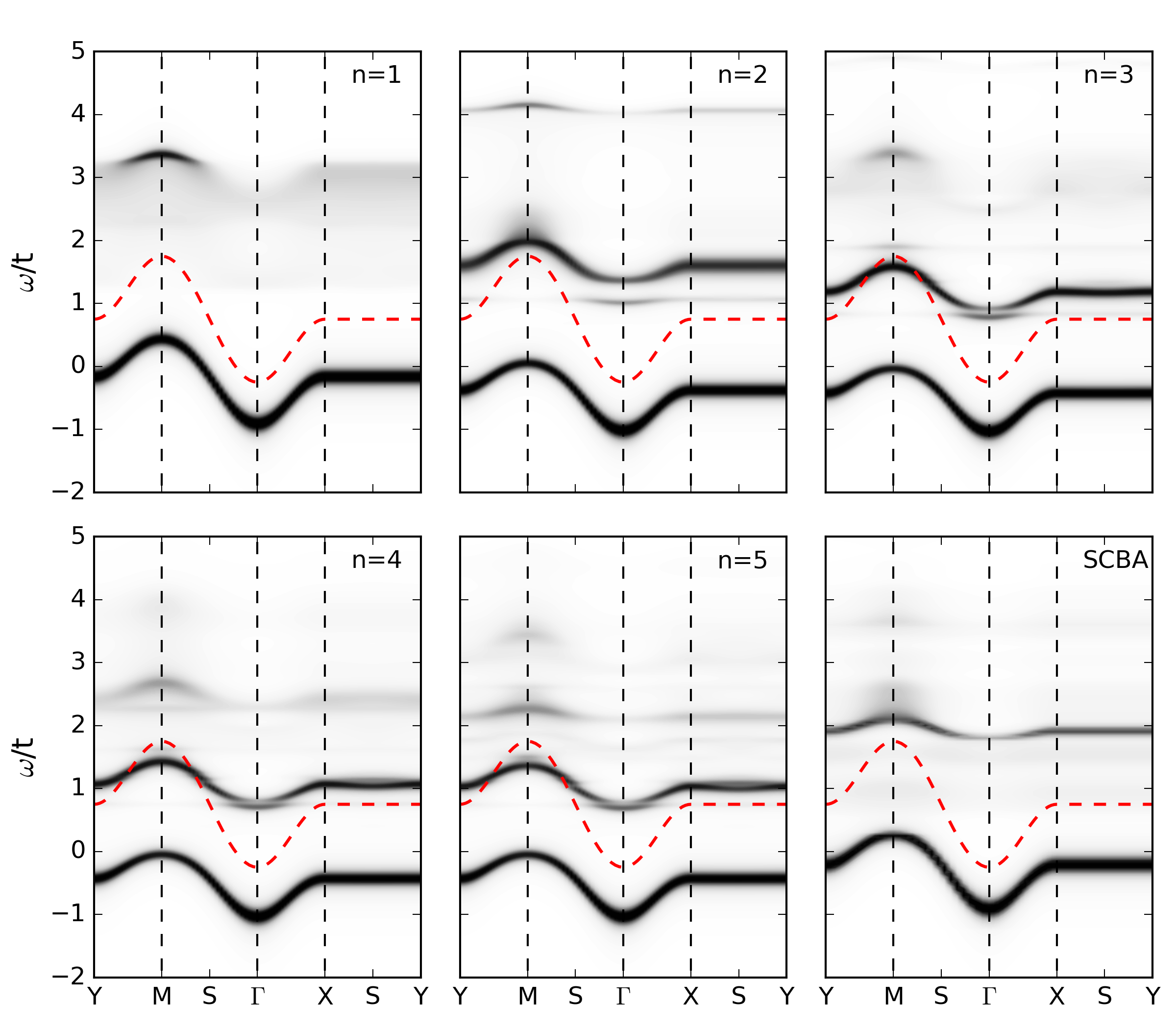}
\caption{(Color online) Momentum dependence of the spectral function
for the variational calculation with a maximum of $n=1$ to $n=5$
orbitons. The 6$^{\mathrm{th}}$ panel shows the corresponding SCBA results.
Parameters: $t=1$, $J=0.5$.
The dashed red line indicates the free electron dispersion.
The spectral function is plotted on a nonlinear $\tanh$-scale to
emphasize the low-amplitude incoherent part of the spectrum.
The high-symmetry points are: $\Gamma=(0,0)$, $X=(\pi,0)$,
$Y=(0,\pi)$, $S=(\pi/2,\pi/2)$, and $M=(\pi,\pi)$.}
\label{fig:VA05}
\end{figure}

Fig.~\ref{fig:VA05}
shows $A(\mathbf{k},\omega)$ along high symmetry cuts in the Brillouin
zone for $n=1$ to $n=5$; the SCBA results are shown in the sixth panel.
Note that we cannot show meaningful VCA results for this large $J/t$
value, for reasons explained above.
The spectrum consists of a low-energy QP band which has most of the
weight, and a broad but low-weight continuum at higher energies.
Focusing first on the QP band, we see that with increasing $n$ it
moves to lower energies and its bandwidth decreases. This is standard
polaronic behavior: a bigger cloud results in a more stable but also
heavier polaron. However, we also see that the QP band is already
converged for $n\approx 2$, in other words the QP binds very few
orbitons in its cloud.

This is expected because for this relatively large $J$, orbiton
excitations are expensive. Moreover, the variational approach properly
implements the constraint of allowing only up to one orbiton per site.
Thus, a cloud with many orbitons would be spread out over many sites.
However, the electron can only interact with orbitons on its NN sites,
so orbitons that are far from it cost (orbital) exchange energy but do
not contribute to a lowering of total energy through carrier-orbiton
interactions. As such, their presence is energetically unfavorable.

Another way to think about it is in terms of an ``effective coupling''.
For lattice polarons, this is defined as $\lambda \sim g^2/(\Omega D)$,
where $g$ is the (suitably averaged) electron-phonon coupling strength,
$\Omega$ is the optical phonon energy and $D$ is the free-electron
bandwidth. For small $\lambda$ the polaron cloud has very few phonons
and the QP properties are hardly changed compared to those of the free
carrier. For large $\lambda$, on the other hand, a so-called
``small polaron'' forms. It has a large boson cloud, with many phonons,
and consequently is a heavy QP.

\begin{figure}[t!]
  \centering
  \includegraphics[width=\columnwidth]{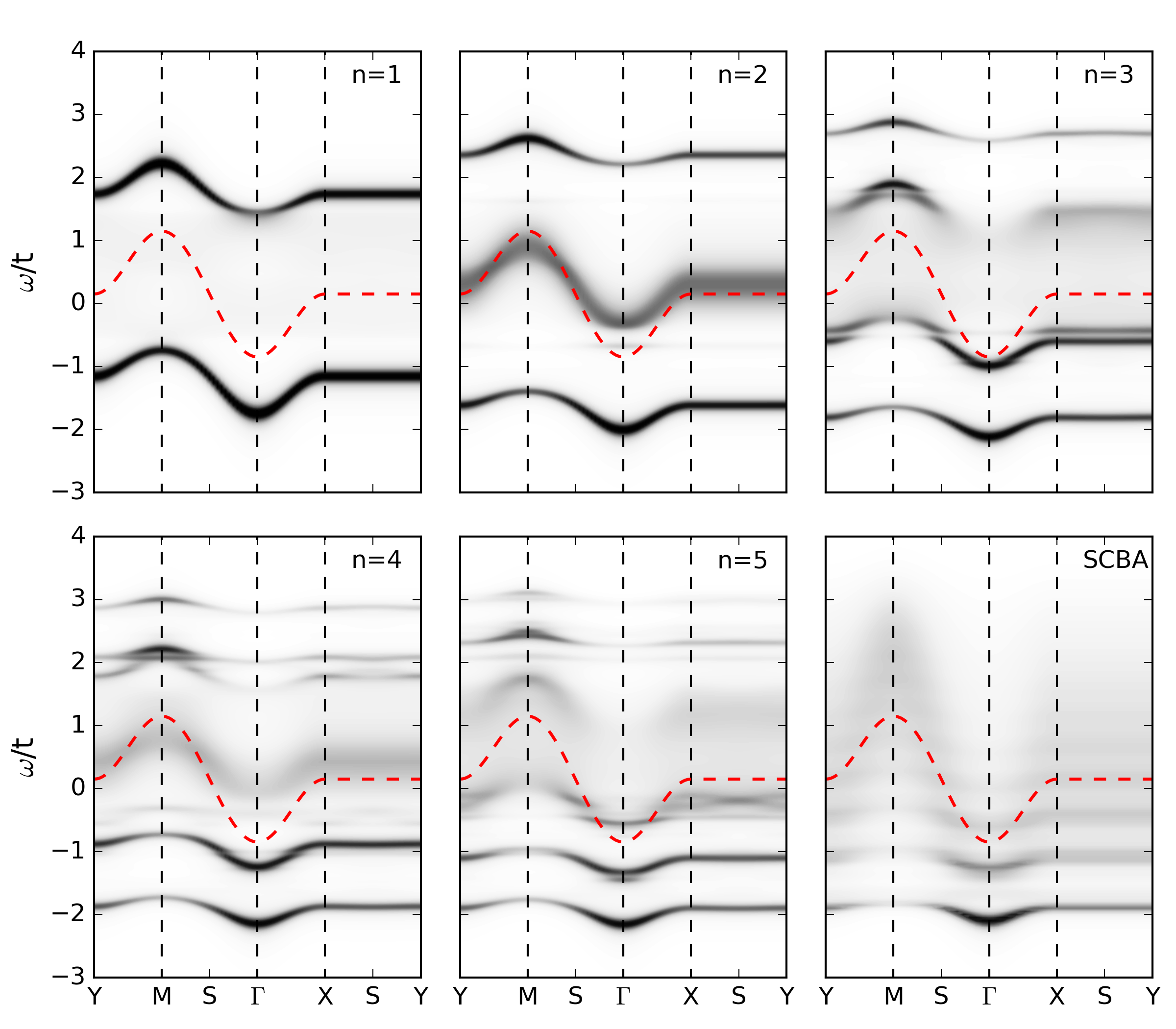}
\caption{(Color online) Same as in Fig.~\ref{fig:VA05} for $J/t=0.1$.
The variational result for $n=6$ is shown in Fig.~\ref{fig:spectral}.
The high-symmetry points are as in Fig.~\ref{fig:VA05}.}
\label{fig:VA01}
\end{figure}

For our problem, $g \sim t, D \sim t, \Omega \sim J$; as a result a
large $J/t$ implies a small $\lambda$ and \emph{vice versa}. The results of
Fig.~\ref{fig:VA05} are, therefore, consistent with this polaron
paradigm of having a small cloud and weakly renormalized QP at small
$\lambda$. Moreover, we also see fairly good agreement with SCBA,
which is also known to be true for small $\lambda$ Holstein lattice
polarons \cite{Ber06,Goo06}. This is easy to understand: for a cloud
consisting of very few bosons, the probability that the carrier
absorbs the last emitted boson (which is the only process allowed in
SCBA) as opposed to any other boson, is of order 1. It is only for
clouds with many bosons that this probability becomes very small. In
this latter case, SCBA is expected to become very inaccurate (as, indeed,
is the case for large $\lambda$ Holstein polarons \cite{Ber06,Goo06})
unless there is some other physical reason to strongly favor the
absorption of the last emitted boson.

Regarding the higher energy continuum, we see that it keeps on
changing up to higher $n$ values, although it converges for $n\approx 5$
(this is why we do not show here the $n=6$ result). This is also expected,
because configurations with many orbitons are energetically expensive
and  contribute primarily to the higher-energy excited states. To
properly describe these high-energy states, therefore, one needs to
consider bigger variational spaces. The agreement between the
variational approach and SCBA is much poorer for this higher energy
continuum; we expect the latter to be less accurate here for reasons
discussed in the previous paragraph.

\begin{figure}[t!]
  \centering
  \includegraphics[width=\columnwidth]{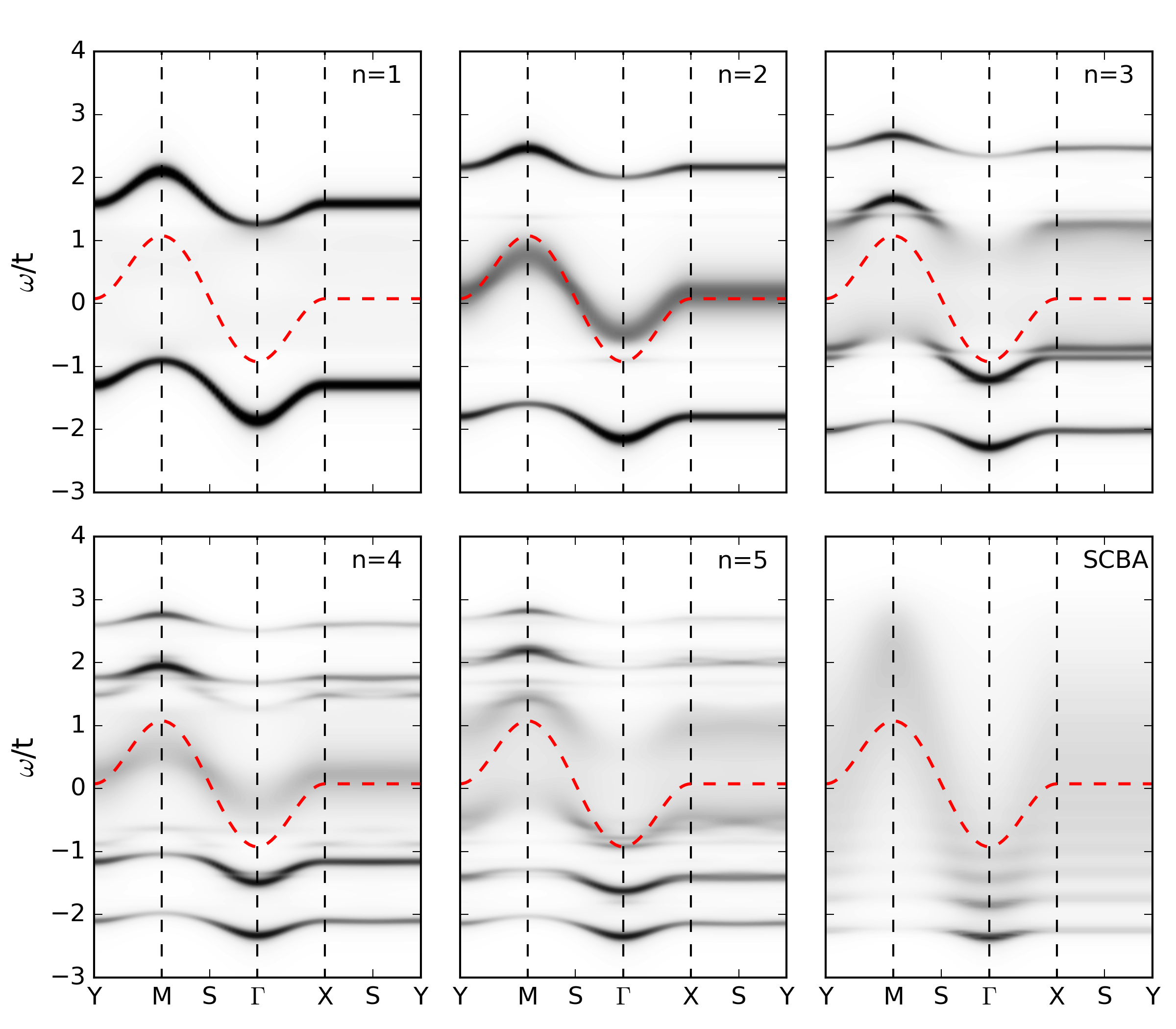}
\caption{(Color online) Same as in Fig.~\ref{fig:VA05} for $J/t=0.05$.
The variational result for $n=6$ is shown in Fig.~\ref{fig:spectral}.
The high-symmetry points are as in Fig.~\ref{fig:VA05}.}
\label{fig:VA005}
\end{figure}

\begin{figure}[b!]
  \centering
  \includegraphics[width=\columnwidth]{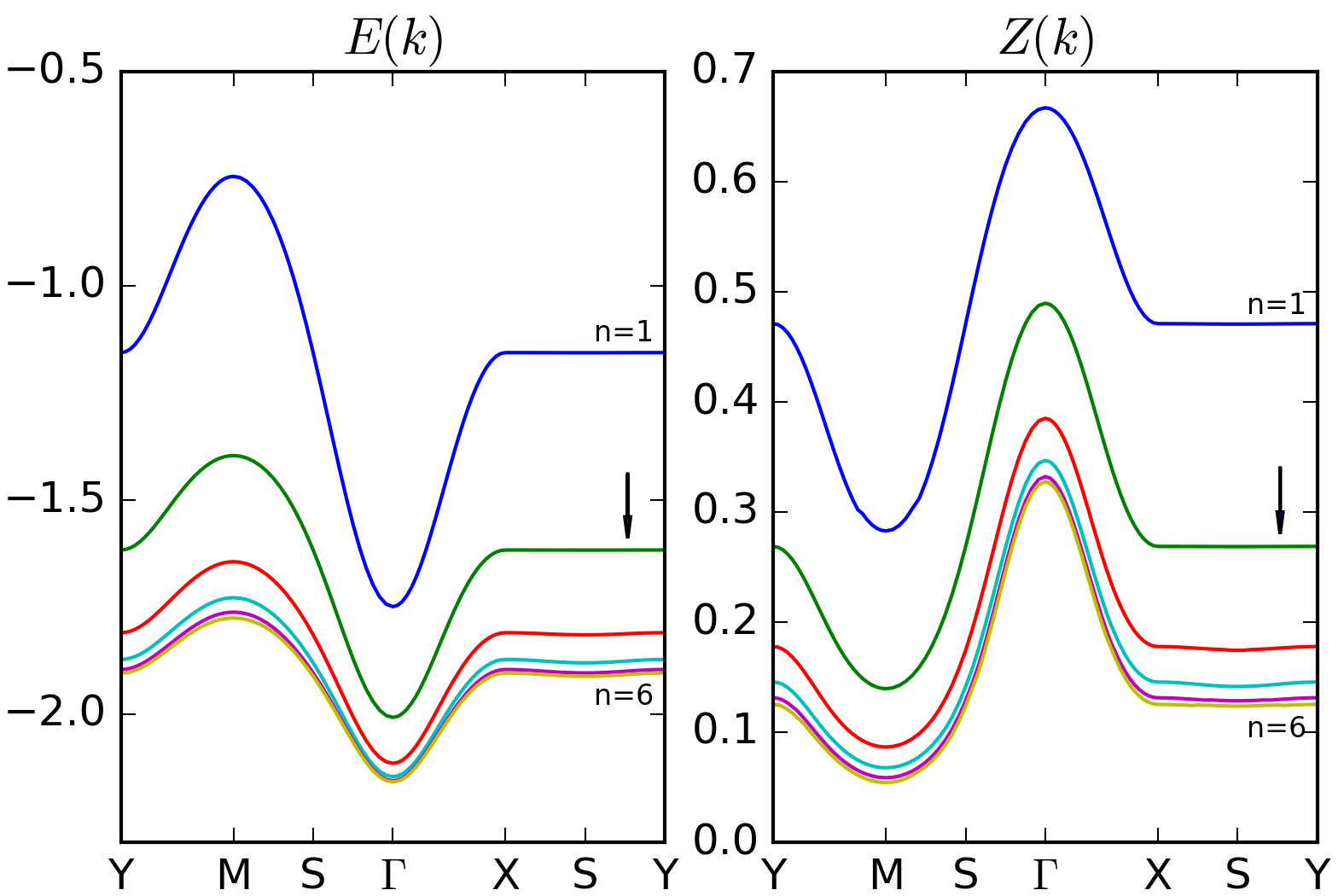}
\caption{(Color online) Ground state energy $E(\mathbf{k})$ and
spectral weight $Z(\mathbf{k})$ calculated numerically for $J=0.1$
(see Fig.~\ref{fig:VA01}) and a maximum of $n=1$ to $n=6$ orbitons
from top to bottom as indicated by arrows.
The high-symmetry points are as in Fig.~\ref{fig:VA05}.}
  \label{fig:converge}
\end{figure}

We have verified that increasing $J/t$ more, \emph{i.e.}, moving toward
even smaller effective couplings, further confirms all the inferences
we have made so far. Clearly, the more interesting question is what
happens in the opposite limit, when the effective coupling increases.
This is shown in Figs.~\ref{fig:VA01} and \ref{fig:VA005} for $J/t=0.1$
and 0.05, respectively. As expected, as we move towards
stronger effective coupling, the QP band reaches convergence more
slowly (at $n\approx 5$ in these cases), moves to even lower energies
and becomes narrower.
For a more direct verification of this see Fig.~\ref{fig:converge}
showing the evolution of QP energy $E(\mathbf{k})$ and spectral weight
$Z(\mathbf{k})$ for increasing number of orbitons at $J=0.1$. One can
clearly see the monotonic convergence of the ground state results, as
should indeed be the case for the variational method. Both of the curves
are practically indistinguishable from one another for the cases $n=5$
and $n=6$, indicating that the result is practically converged for
$n\approx 5$.

The higher energy continuum is also more structured, but it is probably
not yet fully converged. On the
low-energy side, it exhibits ``copies'' of the QP band. The agreement
with SCBA for the QP band is quite good for $J/t=0.1$, but for
$J/t=0.05$ SCBA predicts a much heavier polaron than the variational
approximation (see also Sec. \ref{sec:compare} below).
This is likely due to the fact that SCBA does not
enforce the constraint of allowing only up to one orbiton per site. As
a result, at strong couplings it is likely to overestimate the number
of orbitons in the QP cloud, and therefore predict a heavier QP.

\subsection{Comparison with VCA}
\label{sec:compare}

\begin{figure}[b!]
  \centering
  \includegraphics[width=\columnwidth]{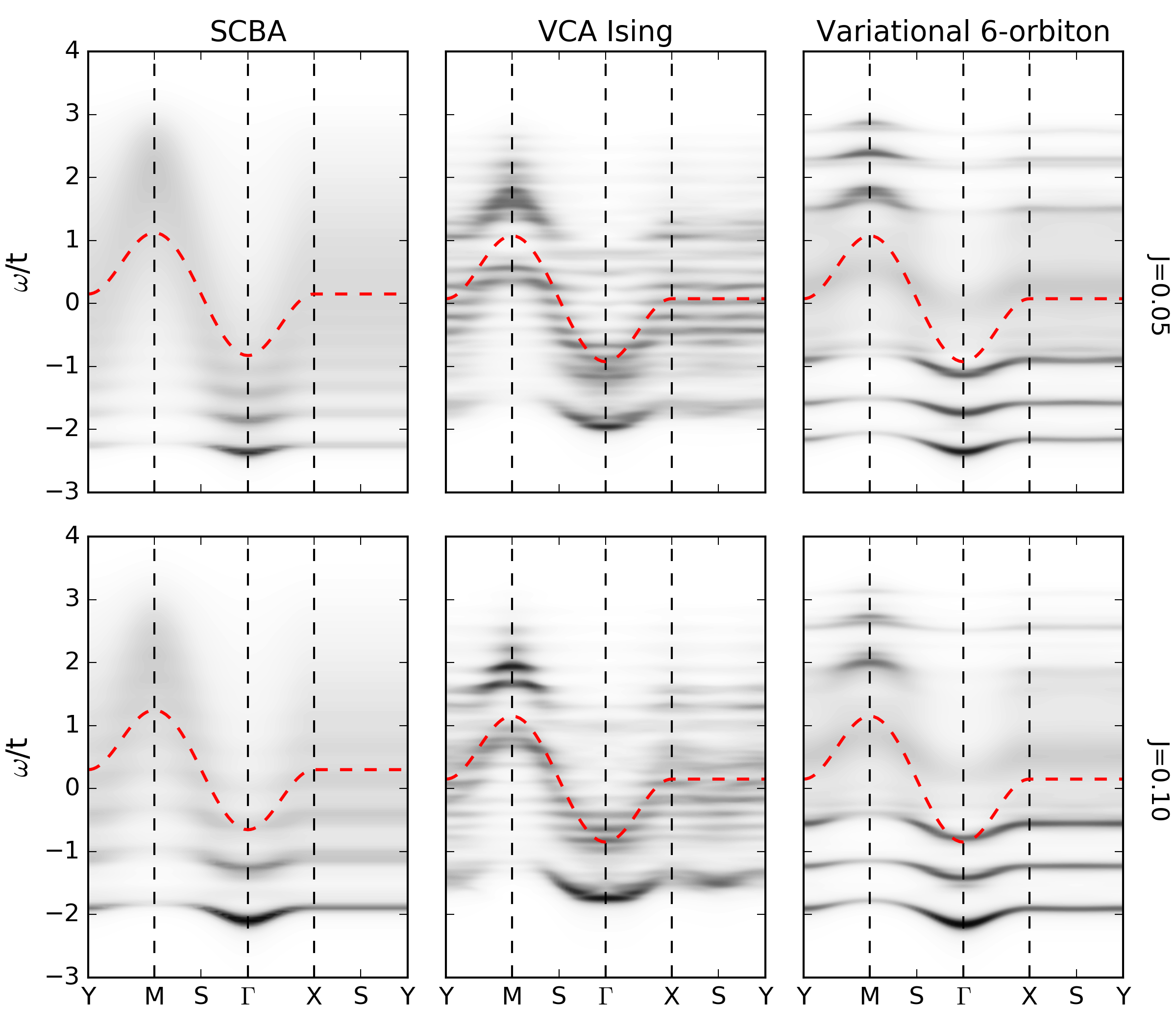}
\caption{(Color online) Comparison of the spectral weights predicted by
SCBA (left), VCA (center) and the variational approach with $n=6$, for
$J/t=0.05$ (top panels) and 0.1 (bottom panels).
The high-symmetry points are as in Fig.~\ref{fig:VA05}.}
\label{fig:spectral}
\end{figure}

Figure~\ref{fig:spectral} presents the comparison of the spectral
function for $J/t=0.05$ and 0.1, for SCBA, VCA and the variational
approach with $n=6$ orbitons. On the first glance, the QP band in the
VCA solution looks more similar to the variational result, but as shown
below the SCBA gives also quite satisfactory results for the QPs.
Although all the solutions exhibit a similar shape, with a lot of
spectral weight around the minimum at the $\Gamma=(0,0)$ point and a
substantial drop in weight with additional
flattening of the spectrum around the $M=(\pi,\pi)$ point, those
features are much more pronounced in the SCBA than in the other two
methods.

Above the QP band, both SCBA and the variational approach show the
emergence of a ladder structure of excited states dissolving quickly
into a broad continuum. Such a ``shake-off'' band also emerges below
the continuum for VCA for $J/t=0.05$, although its weight is rather
small. Thus, it seems reasonable to expect that such a feature does
emerge in the strong coupling limit, although the variational method
and SCBA may overestimate it.

\begin{figure}[t!]
  \centering
  \includegraphics[width=\columnwidth]{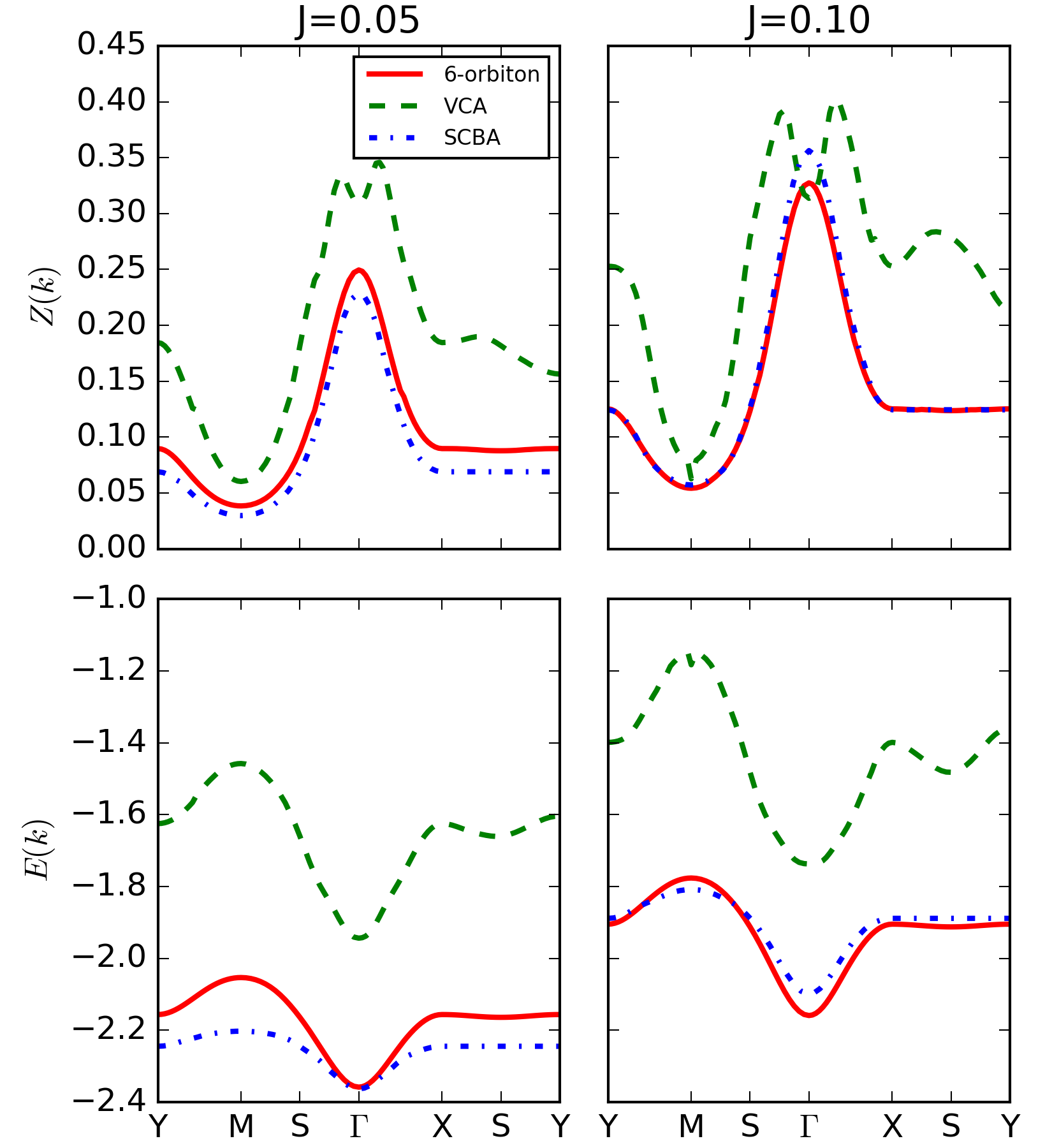}
\caption{(Color online)
Spectral weight $Z(\mathbf{k})$ (top) and ground-state energy
$E(\mathbf{k})$ (bottom) for $J=0.05$
(left) and 0.10 (right) as obtained in the $n=6$ orbiton variational
approach (6-orbiton), in VCA and in SCBA.
The high-symmetry points are as in Fig.~\ref{fig:VA05}.}
\label{fig:weight}
\end{figure}

At even higher energies, SCBA predicts a very featureless continuum,
unlike VCA and the present variational method, which show a more
structured one. However, since neither SCBA nor the variational method
aim to be highly accurate at such high energies, such comparisons are
of limited use. It should also be noted that the apparent dense peak
structure obfuscating the VCA results is not physical but just a
finite-size effect which is the usual byproduct of this method.

Returning to the low-energy polaron, we show its QP spectral weights
$Z(\mathbf{k})$ and ground state energies $E(\mathbf{k})$ in
Fig.~\ref{fig:weight}, obtained by least squares fitting of a
Lorentzian curve to the first peak of the spectrum. As can be seen,
while SCBA and the variational result are in good agreement as to the
spectral weight in both of the presented cases, SCBA always predicts
a heavier QP, especially for the smaller value of $J$,
where it clearly overestimates the binding energy due to not including
the particle constraints, as already discussed. On the other hand, one
can see that the VCA results are in disagreement with both other
results. This can be attributed partly to those calculations being
based on a different model, and so can be affected by a ground state
energy shift, and partly due to the systematic error of the method as
demonstrated by the dense peak structure, which affects the quality of
the peak fit. In any case, the VCA results are in agreement with other
results only qualitatively, although they are usually more similar to
the variational results.

\begin{figure}[t!]
  \centering
  \includegraphics[width=\columnwidth]{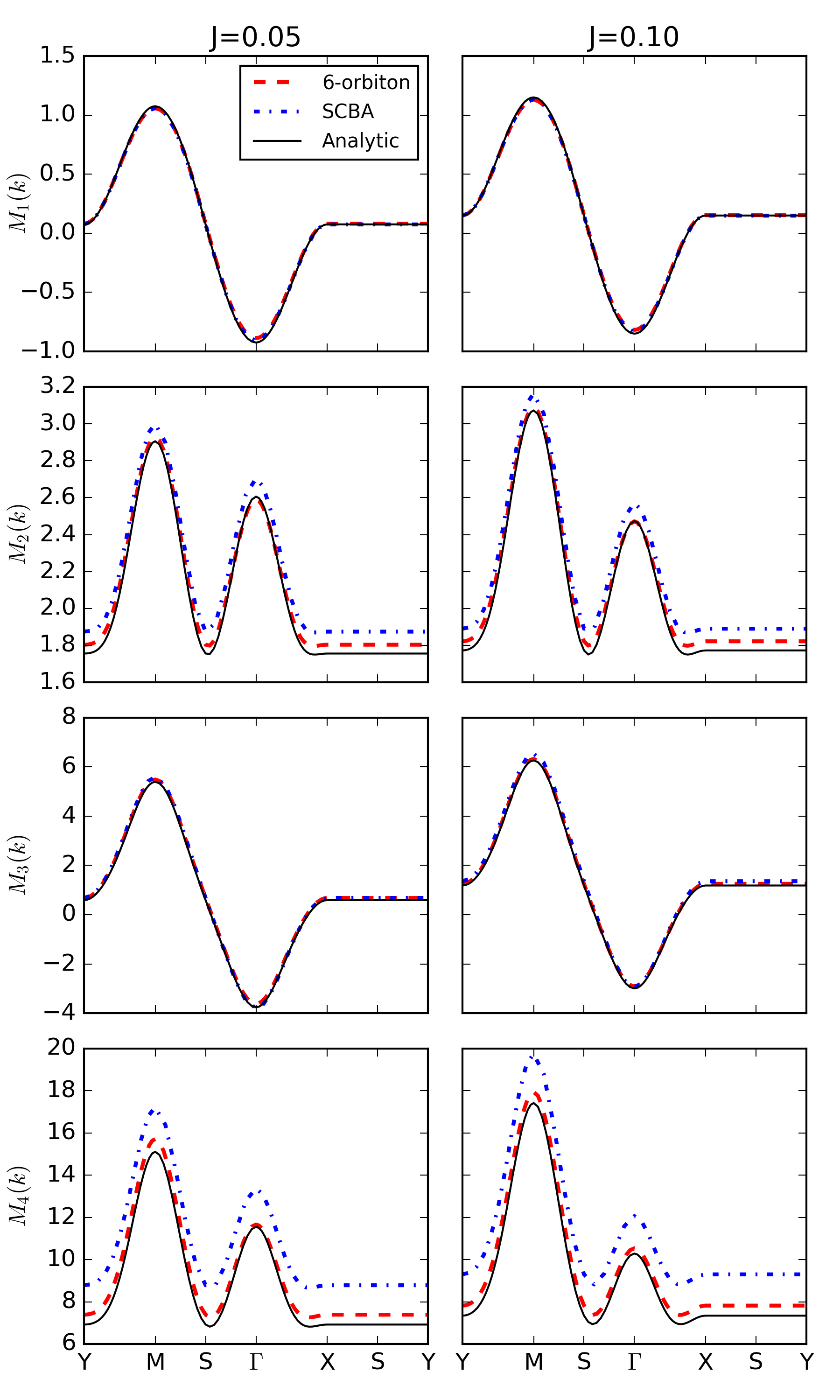}
\caption{(Color online)
First four spectral moments for the calculations performed
in the one-orbiton and two-orbiton variational approximation, and SCBA
as compared to the analytical solution. The high-symmetry points as in
Fig.~\ref{fig:VA05}.}
\label{fig:moments}
\end{figure}

As a final way to compare SCBA and the variational approach in the
strong coupling limit, Fig.~\ref{fig:moments} shows the predicted
first four spectral moments compared to their analytical expressions.
While the agreement is reasonable for both methods up to
$M_3(\mathbf{k})$, the variational approach is clearly more accurate.
Note however, that while agreement with sum rules is obviously
desirable, nevertheless one has to be careful when assessing how
meaningful such an agreement really is \cite{Goo06}.

\subsection{Significance  of inter-orbital hopping}
\label{sec:hopping}

After discussing the similarities and differences between the SCBA
and the variational approach, let us briefly focus on the relative
importance of the $t$ and $J$ energy scales for the QP propagation.
In the strong coupling limit, Fig.~\ref{fig:weight} shows that the
QP bandwidth changes with $r=J/t$, but this dependence is rather weak
for VCA and the variational approach. The bandwidth is reduced from
the free value expected for unhindered inter-orbital hopping, but it
is clearly not proportional to $J$.

\begin{figure}[t!]
  \includegraphics[width=\columnwidth]{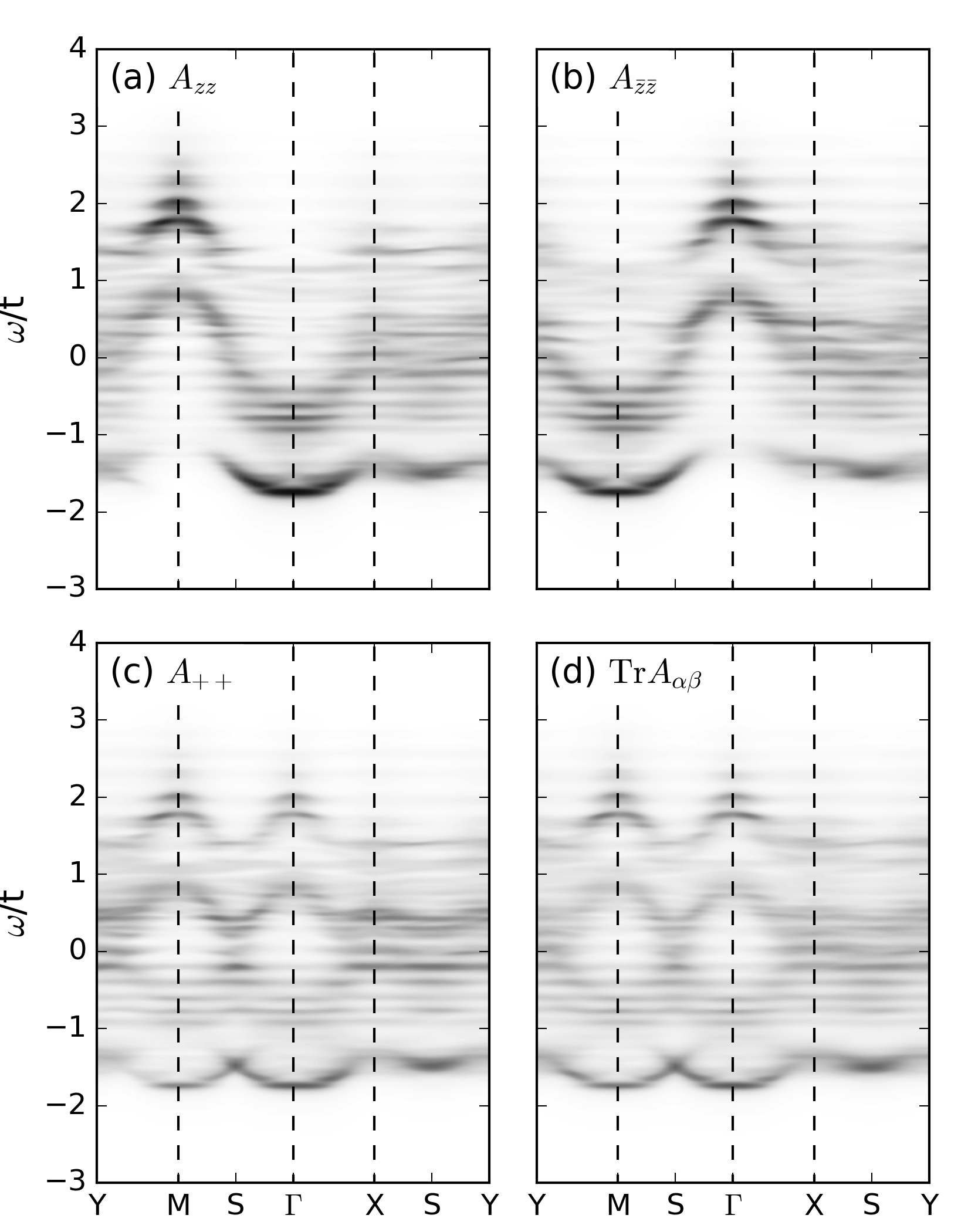}
\caption{
VCA-spectra for AO state at $U=10t$ (corresponding to $J=0.10$) and:
    (a) the $\ket{z}$ orbital,
    (b) the $\ket{\bar{z}}$ orbital,
    (c) the orbital $\ket{+}$, as well as
    (d) the trace over the orbital basis (basis invariant),
    \emph{i.e.}, the normalized sum of (a) and (b).
The high-symmetry points as in Fig.~\ref{fig:VA05}.
\label{fig:AO}}
\end{figure}

We can further substantiate the importance of inter-orbital hopping by
looking at orbitally resolved spectra. To do so, here we show the hole
spectra projected onto the original $\ket{z}$ and $\ket{\bar{z}}$
orbitals. These are given by
\begin{align}
&A_{\alpha\alpha}(\vec{k},\omega) =  \\
&-\frac{1}{\pi} \lim_{\delta\to 0}\Im \left[
  \langle 0| c^{\dagger}_{\vec{k}\alpha}
  \frac{1}{\omega-(\mathcal{H}_t +
  \mathcal{H}_U-E_0)+i\delta}
  c^{}_{\vec{k}\alpha}|0\rangle \right], \nonumber
\end{align}
for $\alpha=z,\bar{z}$ referring to the $3z^2-r^2$ and $x^2-y^2$
orbitals \eqref{basis}. $\mathcal{H}_t + \mathcal{H}_U$ is the
`Hubbard-like' Hamiltonian comprised of Eqs.~(\ref{eq:trans3d}) and
(\ref{eq:hub}), whose half-filled ground state is denoted by
$|0\rangle$ and has energy $E_0$. Spectra for $z$ and $\bar{z}$ are
shown in Figs.~\ref{fig:AO}(a) and~\ref{fig:AO}(b), respectively.

The first notable feature is that the dispersion does not reflect the
doubled unit cell of the AO ground state, as might \emph{a priori} be
expected. This doubling of the unit cell does manifest itself in the
spectrum calculated for the orbital $\ket{\alpha}=\ket{+}$, see
Fig.~\ref{fig:AO}(c), and in the trace shown in Fig.~\ref{fig:AO}(d).
The second relevant aspect is that the dispersion is ``opposite'' for
the $\ket{z}$ and $\ket{\bar{z}}$ orbitals, \emph{i.e.}, the minimum is
either at the $\Gamma$ point (for $\ket{z}$) or at the $M$ point
(for $\ket{\bar{z}}$).

Such an orbital-specific lack of any signature of the unit-cell size
is not expected in a propagation mechanism driven by three-site terms
or orbital flips, and doubling of the unit cell is in fact observed in
the $t_{2g}$ or SU(2) symmetric $t$-$J$ models. However, this feature
arises naturally for a dispersion supported mainly by inter-orbital NN
hopping. The AO ordered ground state consist of alternating $\ket{+}$
and $\ket{-}$ orbitals occupying sublattices $a$ and $b$ respectively.
The interorbital hopping $t_{+-}=t/4$ allows a $\ket{+}$ electron on
sublattice $a$ to move to the $\ket{-}$ orbital of a previously empty
NN site on sublattice $b$ -- the hole has thus moved between NN sites
without disturbing the AO background. The $\ket{z}$ orbital has
positive overlap $+\tfrac{1}{\sqrt{2}}$ with both $\ket{\pm}$ orbitals
and consequently a minimum develops at $\Gamma$ like for a free
particle, see Fig.~\ref{fig:AO}(a). The $\ket{\bar{z}}$ orbital, on the
other hand, has an overlap $\pm\tfrac{1}{\sqrt{2}}$ alternating from
site to site. It translates into the momentum shift moving the minimum
to $M$, as observed in Fig.~\ref{fig:AO}(b).

We have thus identified interorbital hopping as the main driving force
of hole propagation in the $e_g$ model. While the bandwidth is reduced
from its free value through interaction with the AO background, the
SCBA overestimates this effect. This may in turn assign too strong a
role to orbital fluctuations and consequently obscure the origin of the
QP propagation. Finally, we note that an orbital-selective ARPES study
might fail to see the doubling of the unit cell of an orbitally ordered
state, depending on the wave function of the ordered orbital.

\section{Conclusions}
\label{sec:summa}

By analogy with work done for other polaron models, here we
constructed a method to calculate the Green's function for an
orbiton-polaron in a variational space including configurations with
up to six orbitons. We then compared these results with two different
and well established methods, SCBA and VCA. We found that our
variational Green's function converges fast, especially for the
low-energy QP, and that this important feature agrees well with the
corresponding feature obtained in VCA in the strong effective coupling
limit. In contrast, the quasiparticle predicted by SCBA has a
noticeably smaller bandwidth, although general quantitative trends of
the spectra are similar for all three methods.

The higher accuracy of the variational approach can be understood by
noting that its equations of motions allow orbitons to be absorbed
in any order and it consequently includes crossed diagrams, which
are neglected by the SCBA. Moreover, it also implements the local
constraints ignored by SCBA.
In fact, at first sight it is perhaps surprising that the agreement
with SCBA remains as good as it is in the strong effective coupling
limit. This is in contrast to what happens for lattice polarons but
in agreement with what is observed for spin polarons. The common link
with the latter is that like magnons, orbitons are also restricted to
at most one per site. As a result, the cloud is spread over a larger
area and the carrier is more likely to absorb one of the last emitted
orbitons, since they are close to its location (note that at energies
characteristic for the polaron band, which lie well below the
free-particle band, the free carrier cannot propagate). Thus, even
though there is no analogue of the total spin conservation, which
leads to the vanishing of many crossed diagrams and thus improves the
validity of SCBA for spin models, uncrossed diagrams still have higher
probabilities than crossed ones in the present problem. By contrast,
in the small polaron limit of a lattice polaron, the number $n$ of
phonons on the same site can be quite large.
Because they are indistinguishable, the carrier is equally likely to
absorb them in any of $n!$ possible orders,  and SCBA, which only
allows them to be absorbed in the order inverse to that of their
creation, underestimates the contribution of such processes by $1/n!$.

An interesting follow-up question is whether SCBA is still expected to
be qualitatively accurate (even though quantitatively less accurate
than the variational method) for the combined spin-orbiton polarons
\cite{Woh09}, given that it performs reasonably well for either species
of bosons. We expect this to not be the case, because there is no
reason why diagrams with crossed magnon and orbiton lines should be
small compared to the non-crossed ones included in SCBA:
magnons and orbitons can be located on the same site and there is no
conservation law restricting the order in which they can be absorbed.

Moreover, SCBA does not lend itself easily to schemes with multiple
bosonic flavors, which would be the case for the spin-orbital problem,
while the variational method proposed here has already proved highly
accurate for spin-polaron problems (as well as lattice polaron
problems), and can be easily extended to such multi-boson models.
It is thus a promising candidate for investigating carrier propagation
in the full 3D $A$-AF/$C$-AO order in KCuF$_{3}$ which needs to include
both orbiton and spin degrees of freedom, similar to LaMnO$_3$
\cite{Bal01}. Such studies are under way at present.

\begin{acknowledgments}
We thank Krzysztof Wohlfeld for insightful discussions.
We kindly acknowledge support by UBC Stewart Blusson Quantum Matter
Institute,
by Natural Sciences and Engineering Research Council of Canada (NSERC),
by Narodowe Centrum Nauki (NCN, National Science Center)
under Project No.~2012/04/A/ST3/00331, and
by the German Research Foundation (DFG) under grant No. DA 1235.
\end{acknowledgments}

\appendix

\section{Details of the two-orbiton solution}
\label{sec:2orb}

The two-orbiton solution is derived similarly to the one-orbiton case,
but with the orbiton number cutoff set at two. This has no effect on
Eqs.~\eqref{eq:green} to \eqref{eq:G0}, which therefore remain unchanged.
The difference appears first in the equations of motion for the $F_{1}$
propagators, which now also acquire two orbitons propagators on their
right hand side in addition to the terms already listed in
Eqs.~\eqref{eq:f1s} and \eqref{eq:f1bs} indicated here by ``$\ldots$'',
respectively:
\begin{widetext}
  \begin{align}
    \label{eq:f1s2}
    F_{1}^{(2)}(\mathbf{k},\omega,\delta) &=\ldots
    -\frac{t}{4}\sum_{\gamma\epsilon} \left[
      2F_{2}(\mathbf{k},\omega,\epsilon,\gamma)
      -\sqrt{3}\bar{F}_{2}(\mathbf{k},\omega,\epsilon,\gamma)
      e^{i\pi_{y}\cdot\gamma}
    \right] G_{1}(i+\epsilon,i+\delta,\omega),\\
    \label{eq:f1bs2}
    \bar{F}_{1}^{(2)}(\mathbf{k},\omega,\delta) &=\ldots
    -\frac{t}{4}\sum_{\gamma\epsilon}\left[
      2\bar{F}_{2}(\mathbf{k},\omega,\epsilon,\gamma)
      -\sqrt{3}F_{2}(\mathbf{k},\omega,\epsilon,\gamma)
      e^{i\pi_{y}\cdot\gamma} \right]
    G_{1}(i+\epsilon,i+\delta,\omega).
  \end{align}
The generalized Green's functions for states with two orbitons are defined as:
\begin{align}
  \label{eq:f2}
  F_{2}(\mathbf{k},\omega,\epsilon,\gamma) &=
  \bra{\mathbf{k}} \mathcal{G}(\omega)
  \frac{1}{\sqrt{N}}\sum_{i} e^{i\mathbf{k}\cdot\mathbf{R}_{i}}
  f_{i+\epsilon+\gamma}^{\dag} a_{i+\epsilon}^{\dag} a_{i}^{\dag} \ket{0},\\
  \label{eq:f2b}
  \bar{F}_{2}(\mathbf{k},\omega,\epsilon,\gamma) &=
  \bra{\mathbf{k}} \mathcal{G}(\omega)
  \frac{1}{\sqrt{N}}\sum_{i} e^{i(\mathbf{k+Q})\cdot\mathbf{R}_{i}}
  f_{i+\epsilon+\gamma}^{\dag} a_{i+\epsilon}^{\dag} a_{i}^{\dag} \ket{0}.
\end{align}

The equations of motion for these propagators are calculated using
again Dyson's equation. To simplify the analytical solution, we impose
the additional constraint that only configurations with the two
orbitons located on adjacent sites are kept in the variational space.
As shown in Sec. \ref{sec:convergence}, the numerical solution shows that
this is a good approximation for the low-energy quasiparticle. Because
of this and because the orbiton number cutoff is set at two, the
equations of motion for $F_2$ and $\bar{F}_2$ are linked only to
one-orbiton Green's functions:
\begin{align}
  \label{eq:f2s}
  F_{2}(\mathbf{k},\omega,\epsilon,\gamma) = -\frac{t}{4} \sum_{\beta}
    &\left[     (2F_{1}(\mathbf{k},\omega,\epsilon)
    -\sqrt{3}\bar{F}_{1}(\mathbf{k},\omega,\epsilon)e^{i\pi_{y}\cdot\beta})
    G_{2}(i+\epsilon+\beta,i+\epsilon+\gamma)\right.\nonumber\\
    &\left.+(2F_{1}(\mathbf{k},\omega,-\epsilon)
    -\sqrt{3}\bar{F}_{1}(\mathbf{k},\omega,-\epsilon)e^{i\pi_{y}\cdot\beta})
    e^{-i\mathbf{k}\cdot\epsilon}G_{2}(i+\beta,i+\epsilon+\gamma)  \right],\\
  \bar{F}_{2}(\mathbf{k},\omega,\epsilon,\gamma) = -\frac{t}{4} \sum_{\beta}
    &\left[     (2\bar{F}_{1}(\mathbf{k},\omega,\epsilon)
    -\sqrt{3}F_{1}(\mathbf{k},\omega,\epsilon)e^{i\pi_{y}\cdot\beta})
    G_{2}(i+\epsilon+\beta,i+\epsilon+\gamma)\right.\nonumber\\
    &\left.-(2\bar{F}_{1}(\mathbf{k},\omega,-\epsilon)
    -\sqrt{3}F_{1}(\mathbf{k},\omega,-\epsilon)e^{i\pi_{y}\cdot\beta})
    e^{-i\mathbf{k}\cdot\epsilon}G_{2}(i+\beta,i+\epsilon+\gamma)
  \right],
\end{align}
\end{widetext}
where $G_{2}$ functions are real space, non-interacting Green's
functions with spatial constraints imposed by the presence of two
orbitons in the system. They are analogous to the $G_{1}$ functions
defined in equations~\eqref{eq:vi} to~\eqref{eq:gi}, only this time
we need to subtract from the Hamiltonian the terms corresponding to
both sites occupied by the orbitons, namely $\{i, i+\epsilon\}$. The
equation defining these Green's functions is virtually identical to
Eq.~\eqref{eq:gi}, only the range of summation changes. In fact, the
general expression for a constrained Green's function describing a
real-space propagation from site $m$ to site $n$ in a 2D system with
a string of orbitons of length $l$ can be written~as:
\begin{multline}
  \label{eq:gig}
G_{l}(n,m,\omega) =  G_{0}(n,m,\omega-rJ')
+ \sum_{\mean{p,\tilde{p}}}G_{l}(n,p,\omega)\\
\times\left[\frac{t}{4}G_{0}(\tilde{p},m,\omega-rJ')
- 2J'G_{0}(p,m,\omega-rJ')\right],
\end{multline}
where the index $\tilde{p}$ goes over all the sites occupied by an
orbiton and $p$ goes over all the unoccupied sites adjacent to a given
$\tilde{p}$, with the summation going over all such pairs of sites;
$r=4(l+2)$, so in our case $r=16$.

Taken together, these equations define a (bigger) linear system of
inhomogeneous equations, which are then solved to find the solution
in this larger variational space.
A generalization to more orbitons follows along similar lines.
In the case of $n=6$ orbitons, the equations of motions for propagators
$F_n$ with $n\le 6$ are generated and solved numerically.

\bibliographystyle{apsrev4-1}

\end{document}